\documentclass[aps,nofootinbib,showkeys,noshowpacs,preprintnumbers,amsmath,amssymb]{revtex4-2}
\pdfoutput=1 % if your are submitting a pdflatex (i.e. if you have
\usepackage[a4paper, total={6.5in, 9in}]{geometry}
\usepackage[T1]{fontenc} % if needed
\usepackage{microtype}
\usepackage[english]{babel}
\usepackage[utf8]{inputenx}
\usepackage{comment} 
\usepackage{siunitx}
\usepackage{amsmath}
\usepackage{amssymb}
\usepackage{bm}
\usepackage{natbib}
\usepackage{bbold}
\usepackage{epsfig}
\usepackage{hyperref}
\usepackage{graphicx}
\usepackage{csquotes}
\usepackage{mathtools}
\usepackage{slashed}
\usepackage{color,colordvi}
\usepackage{soul}
\usepackage{xcolor}
\usepackage{booktabs,multirow,array}
\usepackage{multirow}
\usepackage{tabularx,ragged2e,booktabs,caption}
\usepackage{comment}
\usepackage{slashed}
\usepackage{float}
\usepackage{caption}
\usepackage{subcaption}
\usepackage[normalem]{ulem}
\renewcommand{\(}{\left(}
\renewcommand{\)}{\right)}
\renewcommand{\{}{\left\lbrace}
\renewcommand{\}}{\right\rbrace}

\newcommand{\order}[1]{\mathcal{O}\({#1}\)}

\def \im{\textrm{Im}}

\newcommand{\msbar}{\overline{\text{MS}}}
\newcommand{\as}{\alpha_\mathrm{s}}
\newcommand{\mh}{M_H}

\newcommand{\pow}[2]{#1^#2}

\newcommand{\gfermi}{G_\mathrm{F}}
\newcommand{\ordas}[1]{\mathcal{O}\left(\alpha_s^{#1}\right)}
\newcommand{\GeV}{\,\mathrm{GeV}}

\newcommand{\MeV}{\,\mathrm{MeV}}

\newcommand{\hs}{\hspace{.4mm}}
\newcommand{\bs}{\hspace{1cm}}

\newcommand{\mtsq}{M_{\tau}^2}

\begin{document}
\title{Renormalization group summation and analytic continuation from spacelike to timeline regions.}

%% %simple case: 2 authors, same institution
%% \author{A. Uthor}
%% \author{and A. Nother Author}
%% \affiliation{Institution,\\Address, Country}

% more complex case: 4 authors, 3 institutions, 2 footnotes

\begin{abstract}
Analytic continuation of the perturbative series from spacelike to timelike regions is performed using renormalization group summed perturbation theory~(RGSPT). This method provides an all-order summation of kinematic ``$\pi^2$-terms'' accessible from a given order of a perturbative series. The impact of the summation of these terms is studied for Higgs boson decay and electromagnetic R-ratio in the perturbative QCD. Results obtained using RGSPT have improved convergence behavior in addition to significantly reduced renormalization scale dependence compared to fixed-order perturbation theory (FOPT). The higher-order behavior using the Pad\'e approximant is also studied for processes considered.
\end{abstract}

\author{M. S. A. Alam Khan}
\email{alam.khan1909@gmail.com}
    \affiliation{Centre for High Energy Physics, Indian Institute of Science, Bangalore 560 012, India}
\maketitle

\section{Introduction}
Quantum chromodynamics~(QCD) is the theory of strong interaction describing the interaction among quarks and gluons. It has asymptotic freedom and is perturbative for large momentum transfers. The strong coupling constant ($\as$), which measures the strength of the interaction, has a relatively large value ($\as(2\GeV)\sim 0.3$) compared to the electromagnetic coupling constant ($\alpha=1/137$) for a few $\GeV$s. This large value of $\as$ also introduces certain issues to the perturbation series in the fixed-order perturbation theory (FOPT) formalism. These key issues are mainly convergence, scheme dependence, and renormalization scale dependence. The uncertainties from these sources might constitute a large portion of the total uncertainty in the theoretical predictions. Various perturbative schemes are devised and used in the literature that addresses these issues. They have been applied in various precision determinations of the parameters of the strong interaction and electroweak parameters of the Standard Model (SM). The perturbative renormalization group (RG) plays a key role in improving the fixed-order perturbation series, and various alternative schemes are also used in the literature. Some of these commonly used schemes are the principle of maximal conformality\footnote{For the continuing studies of the theoretical ambiguities of this approach see e.g. the  existing and recent works of Refs.~\cite{Chawdhry:2019uuv,Goriachuk:2021ayq,Baikov:2022zvq} and references therein.} (PMC)~\cite{Brodsky:2013vpa}, optimized perturbation theory (OPT)~\cite{Stevenson:1981vj}, method of effective charges (MEC)~\cite{Grunberg:1982fw}, complete renormalization group improvement (CORGI)~\cite{Maxwell:1999dv,Maxwell:2000mm}, RGSPT~\cite{Ahmady:2002fd}, RG optimized perturbation theory (RGOPT)~\cite{Kneur:2010ss}, the principle of observable effective matching~\cite{Chishtie:2020cen} etc. \par
The perturbative QCD (pQCD) calculations to higher orders have been performed in the literature for many physical processes. These calculations mainly use the operator product expansion (OPE) formalism in deep Euclidean spacelike regions where perturbative treatment to QCD is applicable. However, the experimental information is obtained in the timelike regions. The physical quantities, such as R-ratios, are related to the discontinuities of the polarization function of a current correlator across the physical cut. Therefore, the analytic continuation from spacelike to timelike region acts as a bridge in relating the experimental observations with the theoretical predictions. For pQCD, the analytic continuation and associated issues have been pointed out in the early days of QCD, especially in Refs.~\cite{Moorhouse:1976qq,Pennington:1983rz,Pennington:1981cw,Bottino:1980rj,Krasnikov:1982fx,Radyushkin:1982kg}. One of the key issues emerged from these studies is large kinematical $\pi^2-$corrections arising from the imaginary part of the logarithms when analytic continuation is performed. %\tcc{The resulting introduces extra $\pi^2-$terms coming from the discontinuity of the logarithms across the cut.} 
These corrections are significant ($\sim (\pi \beta_0 )^n$) at higher orders of the perturbation theory and found to be dominating the genuine perturbative corrections in Refs.~\cite{Kataev:1995vh,Kataev:2008ym,Kataev:2008ntk,Herzog:2017dtz,Baikov:2005rw,Baikov:2008jh}. There have been numerous attempts to sum these kinematical terms in the literature using the RG in Ref.~\cite{Pivovarov:1991bi}, renormalon motivated naive non-abelianization (NNA) in Ref.~\cite{Broadhurst:2000yc}, using contour improved version of pQCD schemes such as CORGI in Ref.~\cite{Maxwell:2001he}, analytic perturbation theory in Ref.~\cite{Bakulev:2006ex,Bakulev:2010gm} and its variants such as fractional analytic perturbation theory in recent Refs.~\cite{Kotikov:2023meh,Kotikov:2023nvz}. This issue has also been addressed in Ref.~\cite{Ahrens:2008qu,Ahrens:2009cxz} for the Higgs production, for Sudakov logarithms in Ref.~\cite{Magnea:1990zb}, for the pion and nucleon electromagnetic form factors in Ref.~\cite{Bakulev:2000uh}, for deep inelastic scattering and Drell-Yan processes in Ref.~\cite{Parisi:1979xd}, for electromagnetic R-ratio in Refs.~\cite{Nesterenko:2020rbj,Nesterenko:2019rag,Nesterenko:2017wpb} and other in Refs.~\cite{Mueller:2012sma,Ralston:1982pa,Pire:1982iv,Gousset:1994yh}  . For hadronic $\tau$ decays, the  contour improved perturbation theory (CIPT) scheme is used and found to be inconsistent with the OPE expansion. For recent developments, we refer to Refs.~\cite{Hoang:2021nlz,Hoang:2021unk,Benitez-Rathgeb:2021gvw,Benitez-Rathgeb:2022yqb,Benitez-Rathgeb:2022hfj,Gracia:2023qdy,Golterman:2023oml}.\par
This article discusses an alternative analytical approach to sum the kinematical $\pi^2-$terms by performing the analytic continuation using RGSPT. In the RGSPT scheme, the running logarithms accessible from a given order are summed using the RG equation (RGE) in a closed form. These running logarithms are the key ingredients in the analytic continuation in the complex energy plane. This scheme has already been used in the study of $\tau$-decays \cite{Abbas:2012py,Abbas:2013usa,Abbas:2013usa,Caprini:2017ikn,Ananthanarayan:2016kll,Ananthanarayan:2022ufx} and other processes as well in Refs.~\cite{Ahmady:1999xg,Ahmady:2002fd,Ahmady:2002pa,Ananthanarayan:2020umo,Ahmed:2015sna}. In these studies, RGSPT provides better stability with renormalization scale variations. In some cases, it can also improve the convergence of the perturbative series at higher orders compared to the results from the FOPT scheme. We have discussed these issues involving the polarization functions related to the scalar, vector, electromagnetic current, and Higgs self-energies. These correlators are used in various QCD sum rule studies, and improvements provided by RGSPT can be crucial in precisely determining various SM parameters related to QCD and weak interaction physics.\par 
 This article is organized as follows: In
 Section \eqref{sec:intro_RGSPT}, we briefly discuss the summation procedure of RG logarithms in the RGSPT schemes. In section~\eqref{sec:def_RD}, the analytic continuation for the polarization or Adler functions in the FOPT and RGSPT schemes are described. In section~\eqref{sec:applications}, the application of analytic continuation of the perturbative series for various processes in the RGSPT and FOPT are discussed. These processes include the Higgs boson decaying to bottom quark pairs as well as gluon pairs, total hadronic decay width of the Higgs boson, electromagnetic R-ratio, and continuum contribution from the light quarks to muon $g-2$. The summary and conclusion are provided in section~\eqref{sec:summary}. The supplementary material needed for various sections can be found in the appendix~\eqref{app:mass_run}, \eqref{app:Rem}, \eqref{app:D_scalar}, \eqref{app:Wilson}, \eqref{app:D_higgs}.
\section{Review of the RGSPT\label{sec:intro_RGSPT}}
Various observables in the pQCD are calculated in the high energy limit, and the perturbative series is expressed in terms of the $\as$ and masses of the quarks ($m_q$). The observables are independent of the scheme used and variations of the renormalization scale. This results in the cancellation of the renormalization scale dependence among the coefficients from different orders of the perturbation series. Practically, all order coefficients of a perturbative series of a physical process are not available for QCD, and the effects of scale dependence can be seen in the fixed order results. These perturbative series also have issues of poor convergence behavior. These effects can lead to substantial theoretical uncertainty, and various perturbative schemes are used in the literature to bring them under control.\par
 RGSPT is a perturbative scheme where RGE is used to systematically sum the running RG logarithms accessible from a given order of the perturbation theory. The closed-form results are found to be less sensitive to the renormalization scale, and hence we get a significant reduction in the theoretical uncertainties. The running logarithms play a key role in the analytic continuation of a perturbative series from spacelike to timelike regions. Their summation, therefore, is necessary so that the effects of analytic continuation can be controlled by summing them to all orders.   \par
A perturbative series $S(Q^2)$ in pQCD can be written as:
\begin{align}
    \mathcal{S}\equiv \sum_{i=0,j=0}^{n,i} T_{i,j} x^i L^j \,,
    \label{eq:Pseris}
    \end{align}
where $x=\as(\mu^2)/\pi$ and $L=\log(\mu^2/Q^2)$. This series can be rearranged into an RG summed series as follows:
\begin{align}
    \mathcal{S}^\Sigma= \sum_{i=0} x^i S_i (x\hs L)\,,
\end{align}
 where,
 \begin{align}
     S_i (x\hs L)=\sum_{j=0}^{\infty} T_{i+j,j} \left(x \hs L\right)^j\,.
     \label{eq:SRcoef}
 \end{align}
 The RG evolution of the perturbative series in Eq.~\eqref{eq:Pseris} with associated anomalous dimension, $\gamma_S$, is given by:
 \begin{align}
     \mu^2 \frac{d}{d\mu^2} \mathcal{S}(Q^2)=\gamma_S \mathcal{S}(Q^2)\,,
     \label{eq:RGE}
 \end{align}
and $\gamma_S$ is given by:
 \begin{align}
     \gamma_S=\sum_{i=0}\gamma_i x^{i+1}\,.
 \end{align}
 The RGE in Eq.~\eqref{eq:RGE} results in constrain on the summed coefficients, $S_i(x\hs L)$ in Eq.~\eqref{eq:SRcoef}, and the recurrence relation between different coefficients are given by:
\begin{align}
  \left(\sum_{i=0}^{n} \frac{\beta_{i}}{u^{n-i-1}}\frac{d}{d\hs u}\left(u^{n-i} S_{n-i}(u)\right)+\gamma_i S_{n-i}(u)\right)-S'_n(u)=0\,,
\end{align}
 where $u\equiv x \hs L$. The first three solutions to the above recurrence relation are given by:
 \begin{align}
     S_0(u)=&T_{0,0} w^{-\tilde{\gamma }_0}\,,\hs S_1(u)=T_{1,0} w^{-\tilde{\gamma }_0-1}+T_{0,0} w^{-\tilde{\gamma }_0-1} \left(\tilde{\beta }_1 \tilde{\gamma }_0 (w-\log (w)-1)-(w-1) \tilde{\gamma }_1\right)\\
      S_2(u)=&T_{2,0} w^{-\tilde{\gamma }_0-2}-T_{1,0} w^{-\tilde{\gamma }_0-2} \left(\tilde{\beta }_1 \left(\tilde{\gamma }_0 (-w+\log (w)+1)+\log (w)\right)+(w-1) \tilde{\gamma }_1\right)\nonumber\\&+\frac{1}{2} T_{0,0} w^{-\tilde{\gamma }_0-2} \Big\lbrace-\tilde{\beta }_1 \tilde{\gamma }_1 \left(2 (w-1) \tilde{\gamma }_0 (w-\log (w)-1)-w^2+2 \log (w)+1\right)\nonumber\\&\bs\bs\bs+(w-1) \Big((w-1) \tilde{\beta }_2 \tilde{\gamma }_0+(w-1) \tilde{\gamma }_1^2-(w+1) \tilde{\gamma }_2\Big)\nonumber\\&\bs\bs\bs+\tilde{\beta }_1^2 \tilde{\gamma }_0 (w-\log (w)-1) \left(\tilde{\gamma }_0 (w-\log (w)-1)-w-\log (w)+1\right)\Big\rbrace\,,
 \end{align}
 where $w\equiv 1-\beta_0 \hs u$, $\tilde{X}\equiv X/\beta_0$ and $T_{i,0}$ are the coefficients at higher order perturbation theory. The important feature of the above procedure is that the most general term of RGSPT is given by:
 \begin{align}
     \Omega_{n,z}\equiv\frac{\log^n(w)}{w^z}\,,
     \label{eq:coef_rgspt}
 \end{align}
 where $n$ is a positive integer and z is real. This term will be used in the next section for the analytic continuation and resummation of the kinematical $\pi^2-$terms.
 \section{Analytic continuation from spacelike to timelike regions}\label{sec:def_RD}
 The polarization functions ($\Pi(q^2)$) of current correlators for a physical process are calculated using the OPE formalism in the spacelike region where momentum transfer is large and perturbative treatment is applicable. These contributions are calculated by evaluating the Feynman diagrams appearing in a given order of the perturbation theory. Generally, the higher-order calculations are performed in a very special kinematical limit \emph{i.e.} either a small mass, a large mass, or in the expansion in the ratios of the masses of particles depending upon the scales present in theory. The OPE expansion also factorizes the short-distance perturbative part with the long-distance non-perturbative contributions. The long-distance contributions are encoded in terms of the quark and gluon condensates which are determined using the lattice QCD, chiral perturbation theory, optimized perturbation theory, etc. Once the relevant diagrams are evaluated, we get a fixed-order perturbation theory (FOPT) series given in Eq~\eqref{eq:Pseris} as an expansion in $\as$. \par
 In general, $\Pi(q^2)$ is not an RG invariant quantity and does not obey a homogeneous RGE. The Adler functions ($D(Q^2)$) in QCD are the RG invariant quantities derived from the $\Pi(q^2)$ as:
 \begin{align}
     D(Q^2)\equiv -Q^2 \frac{d}{d\hs Q^2}\Pi(Q^2)\,,
     \label{eq:adler}
 \end{align}
where the $\Pi(q^2)$ is calculated at spacelike regions ($Q^2=q^2<0$) and has a cut for the timelike regions ($Q^2=q^2>0$) due to the presence of the logarithms~$\log(\frac{\mu^2}{-q^2})$. The discontinuity across this cut is related to observables that are measured in the experiments in the timelike regions. A systematic study of various processes in the experiments thus requires the theoretical calculations to be valid in the energy regions of interest. It should also have a well-behaved behavior such that a precise determination of the various theoretical parameters can be obtained. For this purpose, proper analytic continuation plays a very important role. This section briefly introduces quantities needed for the analytic continuation and how it is performed using FOPT and RGSPT schemes. These relations include theoretical quantities, such as polarization and Adler functions, and their relation to the experimental quantities, such as R-ratios ($R(s)$) for $e^+e^-$, which are used in the other sections of this article.\par   
 Polarization function $\Pi(Q^2)$ is related to the $R(s)$ by the following dispersion relation:
 \begin{align}
     \Pi(Q^2)=\int_0^{\infty}\frac{R(s)}{(s+Q^2)} d\hs s\,,
 \end{align}
 and the $D(Q^2)$ is obtained from the $\Pi(Q^2)$ as:
 \begin{align}
    D(Q^2)=-Q^2\frac{d}{dQ^2}\Pi(Q^2)=Q^2\int_{0}^{\infty}\frac{R(s)}{(s+Q^2)^2}ds\,.
\end{align}
Theoretical value of the $R(s)$ is obtained from the imaginary part of the $\Pi(Q^2)$~\cite{Poggio:1975af} as:
\begin{equation}
	\begin{aligned}
		R(s)&\equiv\frac{1}{2\pi i }\lim_{\epsilon\rightarrow 0}\left[ \Pi(-s-i \epsilon)-\Pi(-s+i \epsilon)\right]\\&=\frac{1}{2\pi i}\int_{-s+i \epsilon}^{-s-i \epsilon} d\hs q^2 \frac{d}{d\hs q^2}\Pi(q^2)\\&=\frac{-1}{2\pi i} \int_{-s+i \epsilon}^{-s-i \epsilon} \frac{dq^2}{q^2} D(q^2)\\ &=\frac{-1}{2\pi i}\oint_{|x_c|=1} \frac{d x_c}{x_c} D(-x_c s)  
	\end{aligned}
 \end{equation}
	where the contour of the integration does not cross the cut. The RGE plays a key role in the analytic continuation of $\Pi(Q^2)$ or $D(Q^2)$. The running logarithms ($\log(\frac{\mu^2}{-q^2})$) present in them are then analytically continued resulting in the large kinematical ``$\pi^2$" corrections. Once the analytic continuation is achieved, the running logarithms are resumed by setting $\mu^2=s$, leaving behind the large $\pi^2$-terms amplified by $\beta_i$ or $\gamma_i$ coefficients. Such kinematical terms sometimes dominate the genuine perturbative corrections at higher orders starting from N$^3$LO. In some cases, the convergence of the fixed order series is also spoiled by them at lower energies. Such examples for the hadronic Higgs decay width and electromagnetic R-ratio can be found in Ref.~\cite{Herzog:2017dtz}. \par
	The timelike perturbative series $\tilde{\mathcal{S}}(s)$ is calculated from a spacelike series $\mathcal{S}(Q^2)$ from Eq.~\eqref{eq:Pseris} as:
	\begin{align}
		\tilde{\mathcal{S}}(s)&\equiv \lim_{\epsilon\rightarrow0}\frac{1}{2 i}(\mathcal{S}(-s-i \epsilon)-\mathcal{S}(-s+i \epsilon))\nonumber\\
  &=\frac{1}{2\pi i}\int_{-s+i \epsilon}^{-s-i \epsilon}dq^2\frac{d}{dq^2}\mathcal{S}(q^2)\nonumber\\
  &=\frac{-1}{2\pi i}\sum_{i=0} j\hs x^{i}T_{i,j}\int_{-s+i \epsilon}^{-s-i \epsilon} \frac{dq^2}{q^2}\log^{j-1}\left(\frac{\mu^2}{-q^2}\right)\label{eq:ancont_1}\\		&=\frac{-1}{2\pi i}\sum_{i=0} j x^{i}  T_{i,j}\oint_{|x_c|=1} \frac{dx_c}{x_c}\log^{j-1}\left(\frac{\mu^2}{s\hs x_c}\right)\,,\quad \left(\text{substituting }q^2\rightarrow-s \hs x_c \right)\label{eq:ancont_2}\\
  &=\frac{-1}{2\pi}\sum_{i=0} j \hs x^{i}  T_{i,j}\int_{-\pi}^{\pi} d\phi\left(L_s-i \phi\right)^{j-1}\,,\quad \left(\text{substituting} x_c\rightarrow e^{i \phi}\right)\label{eq:ancont_3}\\
  &=\frac{1}{2\pi i}\sum_{i=0} x^{i} T_{i,j}\left( (L_s+i \pi )^{j}-(L_s-i \pi )^{j}\right)\,,
  \label{eq:ancont_FOPT}
	\end{align}
	where $L_s\equiv \log\left(\frac{\mu^2}{s}\right)$. So, analytic continuation for a FOPT series given in Eq.~\eqref{eq:Pseris}, can be obtained by taking the imaginary part by substituting $L\rightarrow L_s\pm i\hs\pi$. The logarithmic terms are resumed by setting $s=\mu^2$, leaving behind only the large $\pi^2-$type corrections.\par
 Given the form of the most general term obtained using RGSPT in Eq.~\eqref{eq:coef_rgspt} where running logarithms are present in the numerator and the denominator, summation of $\pi^2-$terms to all orders is very natural. The analytic continuation for RGSPT can also be directly obtained by substituting $L\rightarrow L_s\pm i\hs  \pi$ as for FOPT. Another form in terms of trigonometric functions can also be derived.\par
	The most general term in Eq.~\eqref{eq:coef_rgspt} for the RGSPT series can be written as:
	\begin{align}
		\frac{\log^m(1-u_1 \log\left(\frac{\mu^2}{-q^2}\right))}{\left(1-u_1 \log\left(\frac{\mu^2}{-q^2}\right)\right)^n}=\partial_{\delta}^m\left(1- u_1\log\left(\frac{\mu^2}{-q^2}\right)\right)^{n-\delta}\Bigg\rvert_{\delta\rightarrow0}
	\end{align}
where $u_1=x\hs \beta_0$ and $m$ is always an integer. Following the steps as in Eqs.~\eqref{eq:ancont_1},\eqref{eq:ancont_2},\eqref{eq:ancont_3}, we get the following result for RGSPT:
\begin{align}
\frac{1}{2\pi i} &\oint_{|q^2|=s} \frac{dq^2}{q^2}\frac{\log^m(1-u_1 \log\left(\frac{\mu^2}{-q^2}\right))}{\left(1-u_1 \log\left(\frac{\mu^2}{-q^2}\right)\right)^n}\nonumber\\&=\lim_{\delta\rightarrow0}\partial_{\delta}^m 
\begin{cases}\frac{\tan ^{-1}\left(\frac{\pi u_1}{1- u_1\hs L_s}\right)}{\pi  u_1}\,, & n=1 \\
\frac{w_s^{-\frac{1}{2} (n-\delta -1)} \sin \left((n-\delta -1) \tan ^{-1}\left(\frac{\pi u_1}{1-u_1\hs  L_s }\right)\right)}{\pi u_1 (n-\delta -1)}\,,& n\neq 1
%	\frac{i \left(1-\beta _0 (L-i \pi ) x\right){}^{\delta -n+1}}{2 \pi  \beta _0 x ( n-\delta-1)}+c.c..
\end{cases}
\label{eq:master_eq}
\end{align}
where, $w_s=\left(1- u_1 \hs L_s  \right)^2+\pi^2 u_1^2$. We can see that all the kinematical $\pi^2$-terms are summed in $w_s$ and $\tan^{-1}(\frac{\pi u_1}{1-u_1 \hs L_s}) $. The large logarithms are also under control as they are always accompanied by $\as$, which is one of the important features of the RGSPT. Hence, both RG improvement, as well as all-order summation is naturally achieved in the RGSPT. In addition, results can be calculated in an analytic form, unlike the analytic QCD methods or numerically evaluating imaginary parts along the contour using CIPT. It should be noted that the $n=1$ case has already been known in the literature~\cite{Groote:2012jq,Groote:2001im} and higher terms, to the best of our knowledge, are new and is the prediction from RGSPT. This is our main result, and its implications and the improvements achieved for various physical processes are discussed in the rest of the article.

\section{Application} \label{sec:applications}
In this section, we discuss the application of summation of the kinematical $\pi^2-$terms and its application in various processes. These processes involve energies ranging from a few $\GeV$s to several hundreds of $\GeV$s. They are calculated from the $\Pi(q^2)$ for the gluon field strength tensor as well as vector, axial-vector, and scalar currents. These polarization functions have applications in hadronic $\tau$ decays, $e^+e^-$ annihilation, and hadronic Higgs decays. For more details about these processes, we refer to a recent review in Ref.~\cite{Pich:2020gzz} and references therein. In addition, some continuum contributions to the experimental values of the charmonium and bottomonium moments~\cite{Kuhn:2007vp,Dehnadi:2011gc,Dehnadi:2015fra,Boito:2019pqp,Boito:2020lyp} as well as in hadronic contribution to muon $(g-2)_\mu$~\cite{Boito:2022rkw,Boito:2022dry} are also worth mentioning. Most of such examples for FOPT are already discussed in the literature~\cite{Herzog:2017dtz}. In this article, we have provided a systematic comparison of FOPT results with the RGSPT. First, we discuss the high-energy processes ($\sim 100\GeV$) relevant to the hadronic Higgs decays, and then we move to vector current processes where intermediate energies ($\sim$ few $\GeV$s) are involved. 

\subsection{Higgs decay in pQCD}
\label{subsec:Hdecay}
\subsubsection{\texorpdfstring{$H\rightarrow \overline{b}b$}{} decay.}
\label{subsec:Hbb}
Higgs decaying to bottom pair is the dominant decay of the Higgs boson and has been of constant interest from both experimental and theoretical points of view. This process is also considered one of the signals in the discovery of the Higgs boson. Theoretically, QCD correction to this process is related to the imaginary part of the $\Pi(q^2)$ of the scalar current correlator. It is known to $\ordas{4}$~\cite{Becchi:1980vz,Broadhurst:1981jk,Chetyrkin:1996sr,Baikov:2005rw,Gorishnii:1990zu,Gorishnii:1991zr,Herzog:2017dtz} and the unknown higher-order coefficients are estimated using the d-Log Pad\'e method in Ref.~\cite{Boito:2021scm}. Recently, a FOPT analysis for this process has been used in Ref.~\cite{Aparisi:2021tym} to calculate the mass of the bottom quark at the scale of Higgs boson mass \emph{i.e.} $m_b(m^2_H)$. It has been found that the FOPT is inapplicable at low energy scale $(\mu\sim m_b)$ due to convergence issues mainly arising from the kinematical $\pi^2-$terms. These shortcomings, however, can be resolved using the RGSPT schemes and discussed in the rest of this subsection. It should be noted that other methods, such as a renormalon motivated large-$\beta_0$ procedure in Ref.~\cite{Broadhurst:2000yc} and analytic QCD approach in Ref.~\cite{Bakulev:2006ex} can also be used to resum the kinematical $\pi^2$-terms. Here, we also show that a significant reduction in the scale uncertainty can also be achieved using RGSPT.\par
The decay of the Higgs to bottom quark pair using pQCD is given by:
\begin{align}
	 \Gamma\left(H\rightarrow \overline{b}b\right) &=\frac{3 G_{F} m_{H}}{4 \sqrt{2} \pi} m^2_{b}(\mu^2) \tilde{\mathcal{S}}(\mu^2)+\text{Other corrections}\,,\\
  &=\Gamma_0 \tilde{\mathcal{S}}(\mu^2)
  \label{eq:hbb}
\end{align}
where $\tilde{\mathcal{S}}(\mu)$ is an analytically continued perturbative series as discussed in Eq.~\eqref{eq:ancont_FOPT}. Other corrections, including electroweak and mixed corrections, are irrelevant to our discussion and therefore ignored in this article.\par
Using the $\ordas{4}$ inputs from Refs.~\cite{Baikov:2005rw,Herzog:2017dtz} and D-log Pad\'e predictions to  $\ordas{8}$ from Ref.~\cite{Boito:2021scm}, the FOPT series has following contributions:
\begin{align}
	\tilde{\mathcal{S}}(m_H^2)=1+ 0.2030+0.0374+ 0.0019 -0.0014-0.0004+6\times10^{-6}&+3\times10^{-5}+5\times10^{-7}\,.
	 	\label{eq:hbbmu_mh}
\end{align}
where $\mu=m_H$ with $\alpha_s(m_H^2)=0.1125$ is taken as inputs using RunDec~\cite{Herren:2017osy}.
However, when the above series is calculated at $\mu=\overline{m}_b=4.18\GeV$ and $\alpha_s(\overline{m}_b^2)=0.2245$, numerical contributions from different terms are found to be:
 \begin{align}
 	\tilde{\mathcal{S}}(\overline{m}_b^2)=&1-0.5659+0.0585+ 0.1469-0.1267+0.0297+ 0.0381 -0.0438+ 0.0114\,,
 	\label{eq:hbbmu_mb}
 \end{align}
and we can see that the effect of the running logarithms and the large kinematical corrections spoil the perturbative nature of the series at the scale $\sim \hs m_b$. Due to the poor convergence behavior of the series in Eq.~\eqref{eq:hbbmu_mb}, only results obtained in Eq.~\eqref{eq:hbbmu_mh} are used in Ref.~\cite{Aparisi:2021tym} to extract the $m_b(m^2_H)$.\par
These shortcomings can be cured using the RGSPT scheme, and it can also improve the convergence compared to FOPT results in Eq.~\eqref{eq:hbbmu_mh} and Eq.~\eqref{eq:hbbmu_mb}. At scale $\mu=m_H$, the RGSPT series has the following contributions:
\begin{align}
		\tilde{\mathcal{S}}^{\small{\Sigma}}(m_H^2)=0.9839+ 0.1894+ 0.0469+ 0.0130+0.0042+ 0.0016+ 0.0007+ 0.0003+0.0002\,.
		\label{eq:hbbmu_mb_sum}
\end{align}
 The perturbative series obtained using  RGSPT is monotonically decreasing, unlike in Eq.~\eqref{eq:hbbmu_mb_sum}), and the truncation uncertainty can also be calculated reliably. When we choose renormalization scale $\mu=m_b$, the perturbation series for RGSPT has the following contributions from different orders:
\begin{align}
	\tilde{\mathcal{S}}^{\tiny{\Sigma}}(m_b^2)=&0.4949+ 0.0427+ 0.0097+0.0033+0.0011+0.0005+ 0.0002+0.0001+ 5\times10^{-5}\,.
\end{align}
Again, the resummation of kinematical $\pi^2-$terms has significantly improved the convergence behavior of the series and it can also be used at such low scales. The RGSPT series to $\order{\alpha_s^4}$ is  monotonically convergent for $\alpha_s\le0.373$, which is around charm mass scale $\mu\sim \overline{m}_c$. While the FOPT series is convergent only for $\alpha_s<0.160$, for $\mu\sim4\overline{m}_b$ for $n_f=5$ active quark flavors. Theoretical uncertainties using FOPT and RGSPT series at scale $\mu=m_H$, when scale is varied in the range $\mu\in\left[m_H/4,2\hs m_H\right]$, have the following numerical values:
\begin{align}
		m^2_b(m_H^2)\tilde{\mathcal{S}}(m_H^2)=9.5655\pm0.0105_{\text{trunc.}}\pm0.0113_{\as}\pm0.0103_{\mu}=9.5655\pm0.0185\\
		m^2_b(m_H^2)\tilde{\mathcal{S}}^{\tiny{\Sigma}}(m_H^2)=9.5384\pm0.0324_{\text{trunc.}}\pm0.0104_{\as}\pm0.0036_{\mu}=9.5384\pm0.0342
\end{align}
where uncertainties are ordered as due to truncation, uncertainties present in the PDG average in $\alpha_s$, and the last one is from the scale variations. We can also rewrite it in the following form:
\begin{align}
    \Gamma^{\text{FOPT}}(H\rightarrow \overline{b}b)&=\Gamma_0 (1.241\pm0.002)\,,\\
    \Gamma^{\text{RGSPT}}(H\rightarrow \overline{b}b)&=\Gamma_0 (1.237\pm0.004)
\end{align}
It should be noted that the truncation uncertainty quoted above FOPT is much smaller than the RGSPT results due to the cancellation between the genuine perturbative contributions and the kinematical $\pi^2-$terms. Hence, the exact nature of these truncation uncertainties directly derived from the FOPT may be misleading and should be dealt with carefully. A safer choice for FOPT would be to estimate such uncertainties from the Adler functions rather than analytically continued series. For the Adler function, we get the following contributions at scale $\mu=m_H$:
\begin{align}
    S(m_H^2)=1.0000+ 0.2030+ 0.0539+ 0.0162+ 0.0058+\cdots\,,
\end{align}
and the last term is $\sim 4.3$ times larger than the N$^4$LO term of Eq.~\eqref{eq:hbbmu_mh}.
Other theoretical uncertainties for RGSPT are significantly smaller compared to FOPT.
\par
\begin{figure}[H]
\centering
	\begin{subfigure}{.49\textwidth}
		\includegraphics[width=\textwidth]{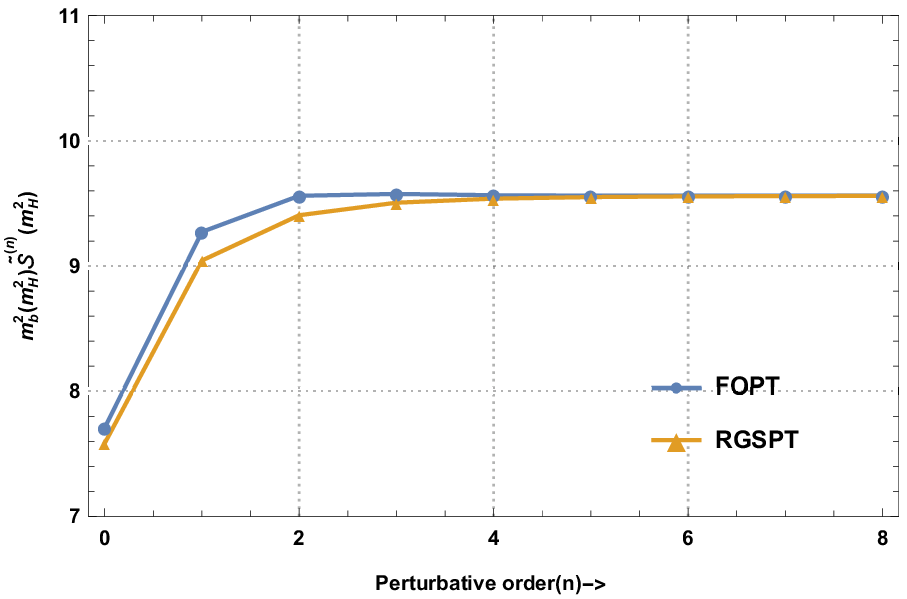}
		\caption{\label{fig:a}}
	\end{subfigure}
	\begin{subfigure}{.49\textwidth}
		\includegraphics[width=\textwidth]{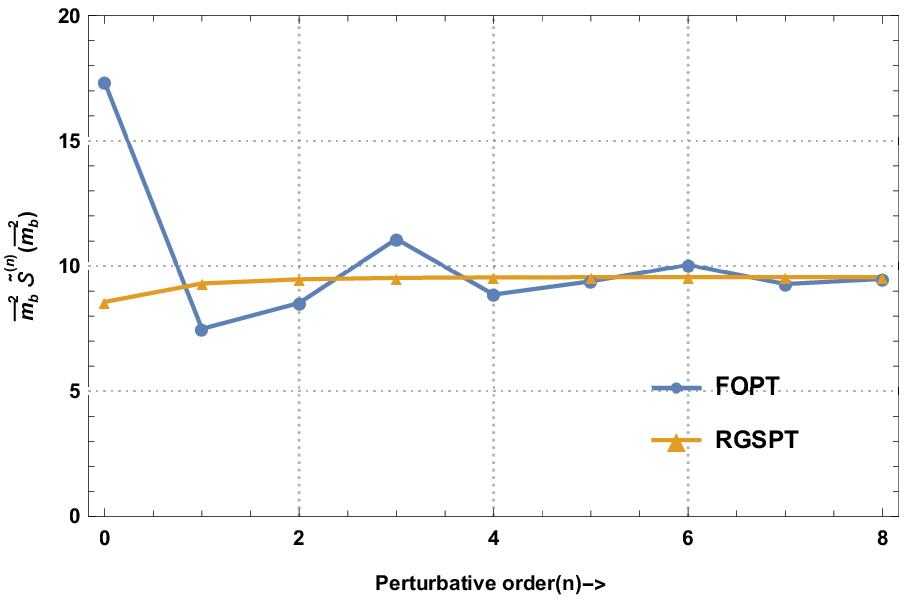}
	\caption{\label{fig:b}}
 \end{subfigure}
	\begin{subfigure}{.49\textwidth}
	\includegraphics[width=\textwidth]{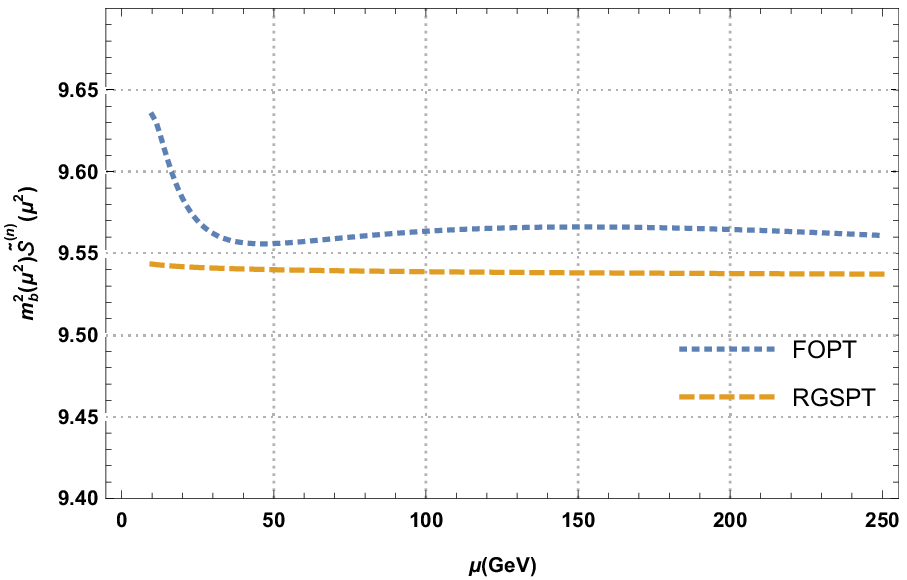}
	\caption{\label{fig:c}}
\end{subfigure}
\caption{\label{fig:Hbb_n} $m_{b}^2(\mu)\delta_{QCD}$ with perturbative order using $\mu=m_H$ in Fig.~\eqref{fig:a}, $\mu=\overline{m}_b$ in Fig.~\eqref{fig:b} and scale dependence for the known $\order{\alpha_s^4}$ terms in the range $\mu\in[10\GeV,2m_H]$ in Fig.~\eqref{fig:c}}.
\end{figure}

 The behavior of $\tilde{\mathcal{S}}_H$ with increasing higher-order contributions for two different scales are presented in Fig.~\eqref{fig:Hbb_n}. The improved convergence over a wider range of renormalization scales using RGSPT allows one to perform the perturbative analysis for a wider energy region compared to FOPT. 
For a relatively large value of the strong coupling at $\mu=m_b$, the convergence of the RGSPT is much better, and it quickly approaches the asymptotic value while the FOPT series oscillates. This can be seen in Fig.~\eqref{fig:b}

These improvements are used in the light-quark mass determination using Borel-Laplace sum rules in Ref.~\cite{AlamKhan:2023ili}.

\subsubsection{\texorpdfstring{$H\rightarrow g g$}{}  decay}\label{subsec:hgg}
This decay mode is the second important channel for the Higgs boson decays. It is mediated by a heavy quark loop and the dominant contribution comes from  a top quark loop due to its large value of the Yukawa coupling, and the decay width is calculated in the heavy-top ($M_H\ll M_t$) limit~\cite{Inami:1982xt}. \par
The decay width of this process in QCD is related to the imaginary part of the self-energy of the Higgs boson ($\Pi^{GG}(q^2)$) via optical theorem. It is given by the following relation~\cite{Inami:1982xt}:
\begin{align}
	\Gamma\left(H\rightarrow gg\right)=\frac{\sqrt{2}G_F}{m_H}|C_1|^2 \im \hs \Pi^{GG}(-m_H^2-i \epsilon)\,,
\end{align}
where $C_1$ is known to N$^4$LO can be found in Refs.~\cite{Chetyrkin:2016uhw,Herzog:2017dtz}. This Wilson coefficient is obtained using the low-energy theorem and decoupling relations for the $\as$. The decoupling relations for the $\as$ and quark masses can also be obtained using RGSPT and provide further RG improvement to this process by reducing the scale dependence~\cite{GAV}.

The perturbative contribution to $\im \hs \Pi^{GG}(q^2)$ are known to $\ordas{4}$~\cite{Kataev:1981gr,Chetyrkin:1997iv,Baikov:2006ch,Schreck:2007um,Herzog:2017dtz} and it can be written as:
\begin{align}
	\im \hs \Pi^{GG}(q^2)\equiv \frac{2 q^4}{\pi}G(x(q),L=0)=\frac{2 q^4}{\pi}\{1+\sum_{i=1}^{\infty}g_i x^i(q)\}
 \label{eq:G2}
\end{align}
 where $x=\alpha^{\left(n_l=5\right)}_s(q^2)/\pi$ and $L= \log(\mu^2/q^2)$. Coefficients $g_i$ can be obtained from Ref.~\cite{Herzog:2017dtz}, and their RG evolution of $G(x,L)$ is obtained by the RG invariance of the quantity:
 \begin{align}
     \mu^2\frac{d}{d\mu^2}\left(\left(\frac{\beta(x)}{x}\right)^2 G(x(\mu), L)\right)=0\,,
     \label{eq:RGEG2}
 \end{align}
where $\beta(x)$ is the QCD beta function defined in Eq.~\eqref{anomalous_dim}. We calculate the RGSPT series for this process by rewriting the perturbative series in terms of the RG summed coefficients, $S_i( x(\mu) L)$, as:
\begin{align}
    G^{\Sigma}(x(\mu),L)=\sum_{i=0}S_i( x(\mu) L) x^i(\mu)\,.
\end{align}
 When the above series is subjected to Eq.~\eqref{eq:RGEG2}, we get the following differential equations among various $S_i( x(\mu) L)$ as:
\begin{align}
	\frac{d S_i(u)}{d u}-\sum _{j=0}^i \frac{\beta _j} {u^{i+j+1}}\frac{d}{d u} \left(u^{i+j+2} S_{i-j}(u)\right)=0\,,
\end{align}
where $u= x(\mu^2)\hs L$. The RGSPT form of the Adler function related to $G(q^2)$ is given by $D_{11}$, and its expression can be found in appendix~\eqref{app:D11_RGSPT}. \par
\begin{figure}[ht]
	\begin{subfigure}{.49\textwidth}
		\includegraphics[width=\textwidth]{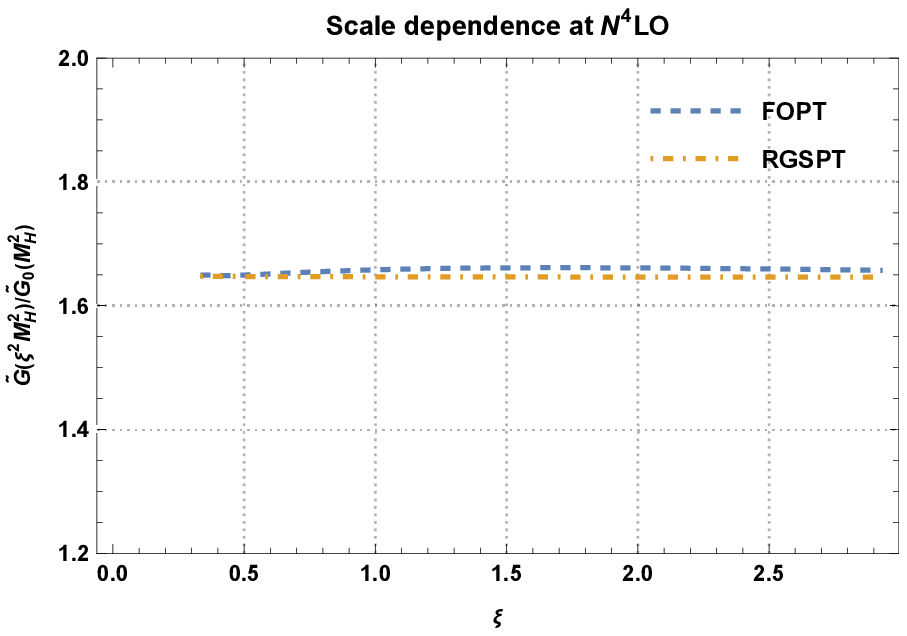}
	\end{subfigure}
	\begin{subfigure}{.49\textwidth}
            \includegraphics[width=\textwidth]{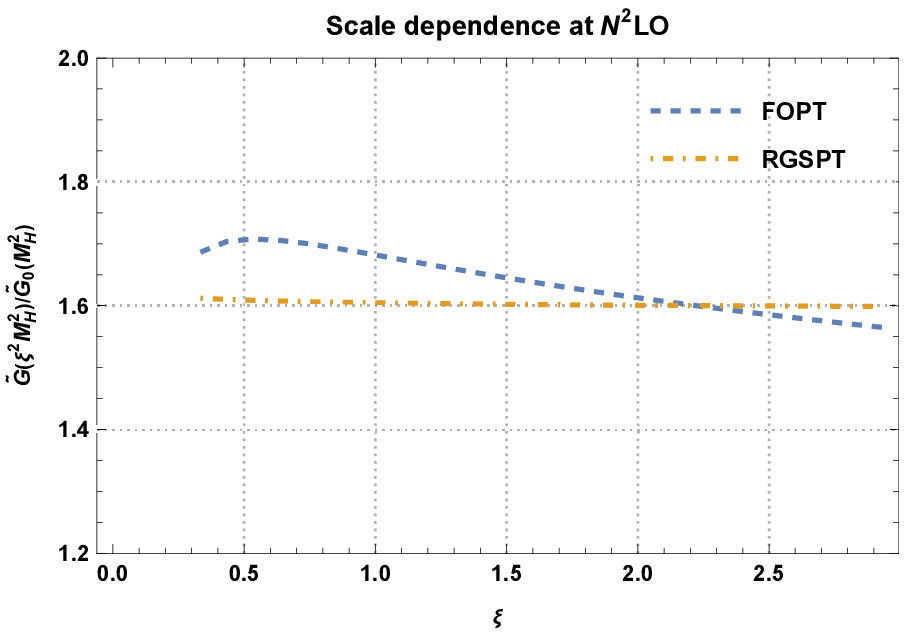}
	\end{subfigure}\newline
 \begin{subfigure}{.49\textwidth}
		\includegraphics[width=\textwidth]{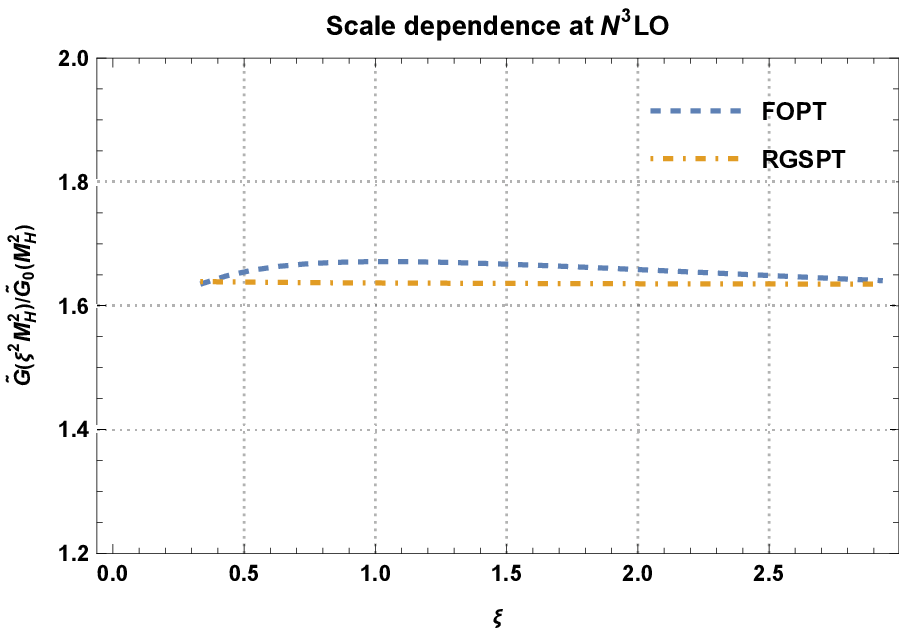}
	\end{subfigure}
	\begin{subfigure}{.49\textwidth}
		\includegraphics[width=\textwidth]{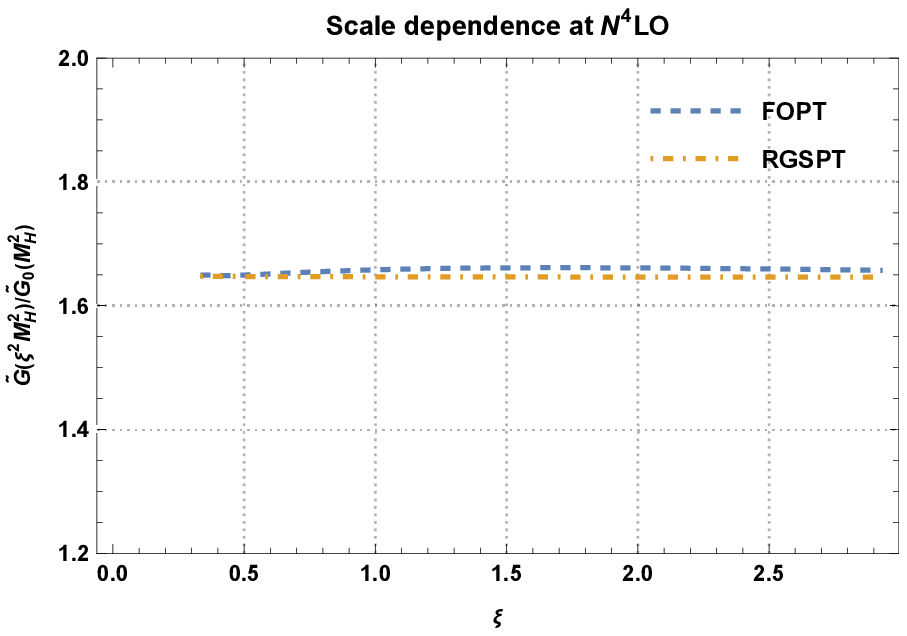}
	\end{subfigure}
 \caption{ \label{fig:G2_scdep}The scale variation of $\tilde{G}(\xi^2 m_H^2)/\tilde{G}_0(m_H^2)$ using known N$^4$LO results in the $\msbar$ scheme.}
\end{figure}

Using $n_l=5$ and setting renormalization scale $\mu=q=m_H$, the $G(x(m_H),0)$ has the following numerical form:
\begin{align}
G^{FOPT}(m^2_H)&=1+3.9523 \alpha _s+6.9555 \alpha _s^2-6.8518 \alpha _s^3-75.2591 \alpha _s^4+\order{\alpha_s^5}\nonumber\\&
=1+ 0.4448+ 0.0881 -0.0098 -0.0121+\cdots\,.
\end{align}
 The N$^4$LO term dominates the N$^3$LO term in the FOPT. In the case of RGSPT, the same quantity is given by:
 \begin{align}
 	G^{RGSPT}(m^2_H)&=0.9555+3.5357 \alpha_s+8.6098\alpha_s^2+20.4939 \alpha_s^3+56.6920 \alpha_s^4+\order{\alpha_s^5}\nonumber\\&
 	=0.9554+0.3979+0.1090+0.0292+ 0.0091+\cdots\,.
 	\label{eq:summedHgg}
 \end{align}
The convergence of the perturbative series is good for RGSPT compared to FOPT. Summation of the kinematical terms also enhances the range of convergence of the perturbation series, and the N$^4$LO term dominates the N$^3$LO term when $\alpha_s(\mu)>0.3$ for $\mu\sim2\GeV$. In addition, a significant reduction in the scale uncertainty can be seen in Fig.~\eqref{fig:G2_scdep}.

 The higher-order behavior and effects of the summation of kinematical terms can be studied by estimating the unknown terms using the Pad\'e approximants. Since only $\order{\alpha_s^4}$ results are known in the literature, we can choose various $W^{\left(n,m\right)}$ Pad\'e approximants for series $S(q^2)$ (before analytic continuation is performed) defined in Eq.~\eqref{eq:Pseris}. The $n$ and $m$ are the exponents of the polynomial in the numerator and denominator of $W^{\left(n,m\right)}$. We obtain the following Pad\'e approximants:
 \begin{align}
     W^{\left(0,4\right)}&=\frac{1}{1- 12.417 x + 49.268 x^2 - 195.212 x^3 - 466.754 x^4}\,,\\
     W^{\left(1,3\right)}&=\frac{1 - 2.391 x}{1 - 14.808 x + 78.957 x^2 - 313.013 x^3}\,,\\
     W^{\left(3,1\right)}&=\frac{1 + 2.571 x - 17.347 x^2 - 146.841 x^3}{1 - 9.846 x}\,.
 \end{align}
The Pad\'e approximant $W^{\left(2,2\right)}$ results in a negative $\order{\alpha_s^5}$ coefficient therefore discarded. The predictions of these approximations are in agreement, and therefore we take their average:
\begin{align}
  \overline{W}=&1+ 12.417 x + 104.905 x^2 + 886.037 x^3 + 8723.76 x^4 +
  89630 x^5 + 906226 x^6 \nonumber\\&\bs+ 8.9855\times10^6 x^7 + 8.87515\times10^7 x^8 +
  8.80468\times10^8 x^9 + 8.76277\times10^9 x^{10}\,,
  \label{eq:hgg_pred}
\end{align}
that can be used for higher-order behavior. An APAP prediction~\cite{Chishtie:1998rz,Ananthanarayan:2020umo} for $\order{x^5}$ coefficient (without analytic continuation) in the large logarithm limit, we obtain:
 \begin{align}
     d_{5}^{APAP}=48056\,,
 \end{align}
 which is nearly half of the predictions from simple Pad\'e approximants in Eq.~\eqref{eq:hgg_pred}. An APAP prediction for the decay width can be found in Ref.~\cite{Abbas:2022wnz}. It will be interesting to compare these predictions with a detailed analysis using the D-log Pad\'e~\cite{Boito:2021scm,Boito:2022rad} or in a large-$\beta_0$ approximation~\cite{Boito:2022fmn}. \par
\begin{figure}[ht]
\centering\includegraphics[width=.7\textwidth]{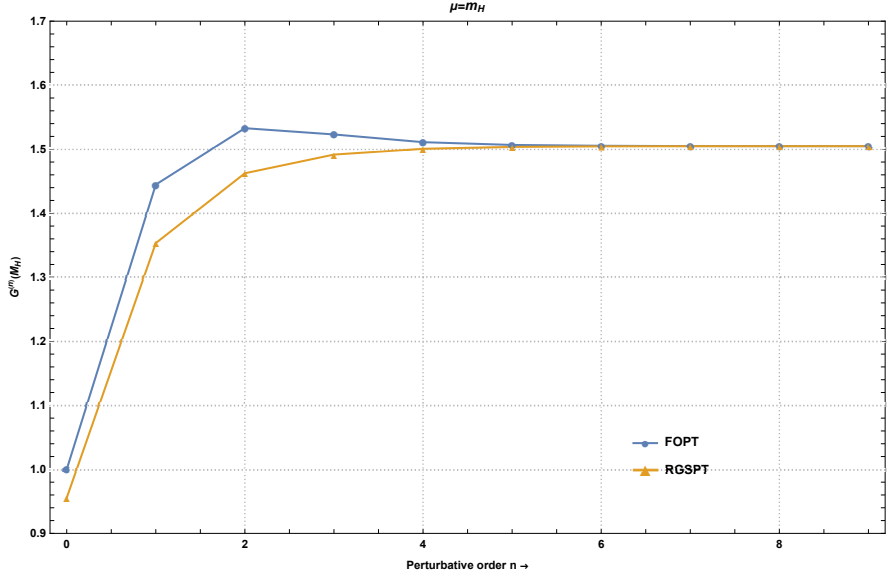}
	\caption{\label{fig:Hgg_n} $G^{(n)}(m_H)$ with perturbative order $n$ in FOPT and RGSPT at scale $\mu=m_H$.}
\end{figure}

 
 Using the above inputs, the higher-order behavior of $H\rightarrow gg$ in FOPT is obtained as:
 \begin{align}
     G^{\text{FOPT}}(m_H^2)=1+0.44477&+ 0.08808 -0.00976 -0.01207 -0.00432 \nonumber\\&-0.00132-0.00051 -0.00015+ 0.00001+\cdots\,,
 \end{align}
 and for RGSPT, the contributions from different order are obtained as:
 \begin{align}
     G^{\text{RGSPT}}(m_H^2)=0.95555&+ 0.39788+ 0.10903+ 0.02921+ 0.00909+ 0.00287\nonumber\\&+ 0.00086+
0.00024+ 0.00006+ 0.00001+\cdots\,,
 \end{align}
 and are shown in Fig.~\eqref{fig:Hgg_n}. It is evident that the 
 RGSPT provides better convergence. The scale dependence from RGSPT and FOPT in different schemes are presented in Fig.~\eqref{fig:G2_scdep}. \par 
  We present our result for this process in different top quark mass schemes used in the literature. The decay width has already been studied very recently in Refs.~\cite{Herzog:2017dtz,Abbas:2022wnz} for scale-invariant (SI), $\overline{\textrm{MS}}$, and on-shell scheme for the top quark mass present in $C_1$. Using top quark mass for SI and $\overline{\textrm{MS}}$ scheme as $\mu_t=m_t(\mu_t)=165\GeV$, and $m_t=173\GeV$ for on shell scheme, we obtain decay width in different schemes using FOPT as:
\begin{align}
\Gamma^{\text{SI,FOPT}}_{\text{N$^4$LO}}	&=\Gamma_0\left(1.8454\pm0.0116_{\text{trunc.}}\pm0.0346_{\alpha_s}\pm0.0159_{\mu}\right)=\Gamma_0\left(1.8454\pm0.0398\right)\,,\nonumber\\
\Gamma^{\overline{\textrm{MS}},\text{FOPT}}_{\text{N$^4$LO}}&=\Gamma_{0}\left(1.8444\pm0.0115_{\text{trunc.}}\pm0.0346_{\alpha_s}\pm0.0135_{\mu}\right)=\Gamma_{0}\left(1.8444\pm0.0389\right)\,,\nonumber\\ \Gamma^{\text{OS,FOPT}}_{\text{N$^4$LO}}&=\Gamma_{0}\left(1.8452\pm0.0111_{\text{trunc.}}\pm0.0346_{\alpha_s}\pm0.0152_{\mu}\right)=\Gamma_{0}\left(1.8452\pm0.0394\right)\,,
\end{align}
where, $\Gamma_{0}\left( H\rightarrow gg\right)=\frac{G_{F}m_H^3}{36\pi^3\sqrt{2}}\alpha_s^2(m^2_H)$ is leading contribution.
The decay width in the RGSPT scheme read:
\begin{align}
\Gamma^{\text{SI,RGSPT}}_{\text{N$^4$LO}}	&=\Gamma_0\left(1.8273\pm0.0177_{\text{\text{trunc.}}}\pm0.0337_{\alpha_s}\pm0.0037_{\mu}\right)=\Gamma_0\left(1.8273\pm0.0382\right)\,,\nonumber\\
\Gamma^{\overline{\textrm{MS}},\text{RGSPT}}_{\text{N$^4$LO}}&=\Gamma_{0}\left(1.8264\pm0.0177_{\text{trunc.}}\pm0.0346_{\alpha_s}\pm0.0067_{\mu}\right)=\Gamma_{0}\left(1.8264\pm0.0386\right)\,,\nonumber\\	\Gamma^{\text{OS,RGSPT}}_{\text{N$^4$LO}}&=\Gamma_{0}\left(1.8271\pm0.0181_{\text{trunc.}}\pm0.0346_{\alpha_s}\pm0.0042_{\mu}\right)=\Gamma_{0}\left(1.8271\pm0.0385\right)\,.
\end{align}
We can also calculate the decay width in the miniMOM (MM) scheme~\cite{vonSmekal:2009ae,Gracey:2013sca} and top quark mass in the on-shell scheme. The results in FOPT and RGSPT schemes are given by:
\begin{align}
\Gamma^{\text{MM,FOPT}}_{\text{N$^4$LO}}&=\Gamma_{0}\left(1.8444\pm0.0111_{\text{trunc.}}\pm0.0346_{\alpha_s}\pm0.0136_{\mu}\right)=\Gamma_{0}\left(1.8444\pm0.0388\right)\,,\nonumber\\	\Gamma^{\text{MM,RGSPT}}_{\text{N$^4$LO}}&=\Gamma_{0}\left(1.8264\pm0.0181_{\text{trunc.}}\pm0.0337_{\alpha_s}\pm0.0047_{\mu}\right)=\Gamma_{0}\left(1.8264\pm0.0385\right)\,.
\end{align}
It should be noted that the scale dependence is calculated by varying scale in the range $\xi\in\left[m_H/3,3 \hs m_H\right]$. It is clear from these results that the $\pi^2-$terms significantly cancel the genuine perturbative corrections, which led to small truncation uncertainty in the FOPT results. A similar behavior is also observed for the $H \rightarrow bb$ process, but here  $\as$ is a primary source of uncertainty.

\subsubsection{Total Hadronic Higgs Decay Width.}
\label{subsec:H_Hadron}
The branching ratio of Higgs boson decays to hadrons is about 70\%~\cite{Davies:2017xsp}. The hadronic decay width of Higgs boson is calculated by constructing an effective Lagrangian from Yukawa term and strong interaction where heavy top quark is integrated out~\cite{Inami:1982xt,Chetyrkin:1996wr,Chetyrkin:1996ke}. The effective Lagrangian has the form:
\begin{align}
\mathcal{L}_\text{eff.}=-\frac{H^0}{v^0}\left(C_1 \left[\mathcal{O}'_1\right]+C_2 \left[\mathcal{O}'_2\right]\right)+\mathcal{L}'\,,
\label{eq:L_eff}
\end{align}
where $H^0$ and $v^0$ are the bare Higgs field and vacuum expectation value. The primed quantities are defined in five-flavored QCD. The coefficients $C_1$ and $C_2$ in Eq.~\eqref{eq:L_eff} are the Wilson coefficients of the operators constructed out of light gluonic and bottom quark degrees of freedom. These Wilson coefficients also carry large logarithms $\sim\log(\mu^2/m_t^2)$, which can be summed using the RGSPT~\cite{GAV}. Since these coefficients do not involve analytic continuation, their FOPT expressions are used in this section. Their numerical expression can be found in the appendix~\eqref{app:Wilson}.
The operators $\mathcal{O}_1^\prime$ and $\mathcal{O}_2^\prime$ are given by:
\begin{align}
    \mathcal{O}_1^\prime&=\left(G_{a,\mu\nu}^{0\prime}\right)^2\,,\\
      \mathcal{O}_2^\prime&=m_b^{0\prime}\overline{b}^{0\prime}b^{0\prime}\,,
\end{align}
where $G^{0\prime}_{a,\mu\nu}$ is the bare gluon field strength, $m_b^{0\prime}$ is the bare bottom quark mass and $b^{0\prime}$ is bare bottom quark field. The current correlators for the operators $\mathcal{O}_1^\prime$ and $\mathcal{O}_2^\prime$ are given by:
\begin{align}
\Pi_{ij}(q^2)=i \int dx \hs e^{i q x} \langle0\vert \mathcal{T}\left[\mathcal{O}^\prime_i,\mathcal{O}^\prime_j\right]\vert 0\rangle\,.
\label{eq:corr}
\end{align}
These correlators can be used to define analytically continued quantities:
\begin{align}
    \Delta_{ii}&=K_{ii}\hs \text{Im}\left(\Pi_{ii}(M_H^2)\right)\,,\\
    \Delta_{12}&=K_{12}\hs \text{Im}\left(\Pi_{12}(M_H^2)+\Pi_{21}(M_H^2)\right)\,,
\end{align}
where $\Pi_{12}(M_H^2)=\Pi_{21}(M_H^2)$, $K_{11}=\left(32\pi M_H^4\right)^{-1}$ and $K_{12}=K_{22}=\left(6\pi M_H^2 m_b^2\right)^{-1}$. The $\Delta_{11}$ is proportional to $G(q^2)$ in Eq.~\eqref{eq:G2} and $\Delta_{22}$ is related to the $\tilde{\mathcal{S}}$ in Eq.~\eqref{eq:hbb}.\par 
Using the quantities defined above, the total hadronic Higgs decay width is given by:
\begin{align}
    \Gamma\left(\text{H}\rightarrow \text{Hadrons}\right)&=A_{b\overline{b}}\left(C^2_2(1+\Delta_{22})+C_1 C_2 \Delta_{12}\right)+A_{gg}C_1^2 \Delta_{11}\,,
    \label{eq:Hhad}
\end{align}
     where,
     \begin{align}
         A_{b\overline{b}}\left(\mu^2\right)=\frac{3}{4\pi \sqrt{2}}\gfermi M_H \hs m_b^2\left(\mu^2\right)\,,\quad
         A_{gg}=\frac{4}{\pi \sqrt{2}}\gfermi \pow{\mh}{3}\,.
     \end{align}\par
\begin{figure}[H]
		\centering
		\includegraphics[width=.7\textwidth]{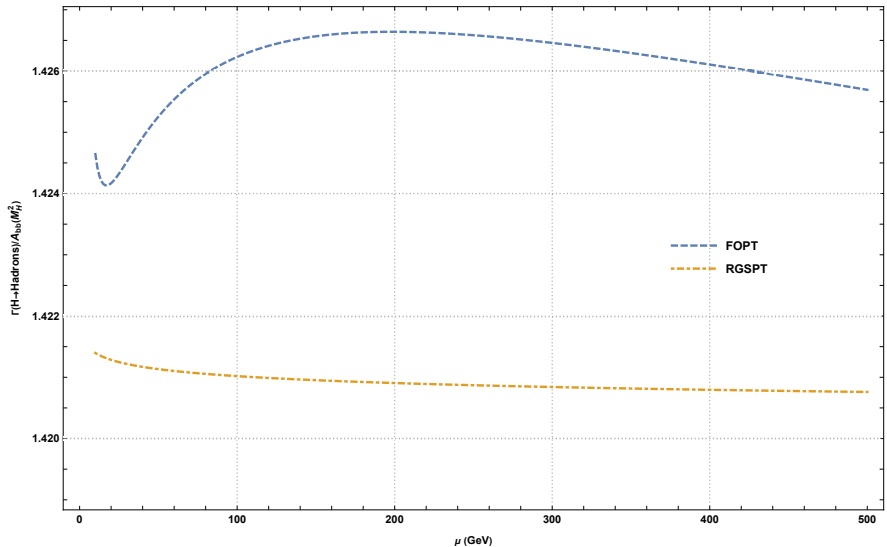}
  \caption{\label{fig:HHad} The scale-dependence of the Hadronic Higgs decay in the $\msbar$ scheme.}
\end{figure}

 Recently, the $\ordas{4}$ corrections to hadronic Higgs decay width are presented in Ref.~\cite{Herzog:2017dtz,Davies:2017xsp} with additional bottom quark mass corrections to $H\rightarrow g g$ in Ref.~\cite{Davies:2017rle}. A PMC analysis for $H\rightarrow g g$ and $H\rightarrow b b$ processes can be found in the Ref.~\cite{Wang:2013bla}.  The total Higgs hadronic decay width is presented using these results in Table~\eqref{tab:Hhad1} and scale-dependence in Fig.~\eqref{fig:HHad}. It should be noted that the theoretical uncertainty in RGSPT is dominated by truncation uncertainty from $\Gamma\left(H\rightarrow \overline{b}b\right)$ contributions. The central value obtained from RGSPT is slightly smaller than FOPT due to the summation of $\pi^2$-terms but agrees within the quoted uncertainty.
 \begin{table}[H]
     \centering
     \begin{tabular}{|c|c|c|c|c|c|c|}
     \hline
          \text{Scheme}&$\Gamma\left(\text{H}\rightarrow \text{Hadrons}\right)(\MeV)$& \multicolumn{5}{c|}{Source of Uncertainty}\\
          \cline{3-7}\text{}&\text{}&\text{Trunc.}& $\delta\as$&$\delta m_b$&$\delta M_H$& $\mu$\\\hline
\text{FOPT}&$2.7082\pm 0.0206$&0.0045&$0.0048$&$0.0183$&$0.0051$&$0.0043$\\\hline
\text{RGSPT}&$2.6978\pm 0.0226$&$0.0115$&$0.0045$&$0.0182$&$0.0051$&$0.0008$\\\hline
     \end{tabular}
     \caption{Total hadronic Higgs decay width in FOPT and RGSPT in the $\msbar-$scheme and the sources of uncertainties. The scale dependence is calculated by varying the renormalization scale in the range $\mu\in\left[10,500\right]\GeV$.}
     \label{tab:Hhad1}
 \end{table} 
 
Now, we move on to the application of analytic continuation of the polarization function involving momentum transfer in the intermediate energy range ($\sim$ few $\GeV$).
\subsection{The electromagnetic R-ratio}
\label{subsec:Rem}
The electromagnetic R-ratio is obtained from the imaginary part of the polarization function for the vector current.  It is an important observable that can be used in the determination of $\as$ using QCD sum rules~\cite{Boito:2018yvl}. The singlet and non-non singlet contributions to process $e^+e^-\rightarrow$ hadrons have been calculated to N$^4$LO in Refs.~\cite{Baikov:2008jh,Baikov:2010je,Baikov:2011pkl,Baikov:2012er,Herzog:2017dtz}. The numerical results presented in Ref.~\cite{Herzog:2017dtz} show that the kinematical $\pi^2$-terms start dominating at N$^3$LO for $n_f=1,2,\cdots6$ flavors of the quark. In this subsection, we study the summation of these terms using RGSPT and their effects, as discussed in the previous sections. The numerical expressions for the Adler functions used in this subsection can be found in the appendix~\eqref{app:Rem}. \par 
The leading-order correction to the Alder function in the massless limit is a constant; therefore, no further enhancement is arising due to the prefactors from the anomalous dimensions. Hence, the effects of analytic continuation are expected to be milder than the $H\rightarrow gg$ or $H\rightarrow bb $ processes in the hadronic decay width. However, these effects can still be enhanced at low energies by a relatively large value of the strong coupling. For RGSPT, both the $\as$ and the kinematical terms are present in the denominators, which also results in further reduction in the truncation uncertainties, as observed in previous subsections. \par
For numerical comparison, we are considering the two different cases. In the first case, we consider momentum transfer below the $J/\psi$ threshold where active quark flavors are $n_f=3$ and $\alpha_s\simeq0.3$. In the second analysis, the momentum transfer is above the mass of the $c\overline{c}$ resonance and below the $\Upsilon$ threshold for which active flavors are $n_f=4$ and $\alpha_s\simeq0.2$. These two cases are already discussed in Ref.~\cite{Herzog:2017dtz} in detail, and we will compare our results with theirs. \par
 \subsubsection{Case-I: Three Active Flavor}\label{subsub:rem3}
\begin{figure}[ht]		
  \centering
				\includegraphics[width=.49\textwidth]{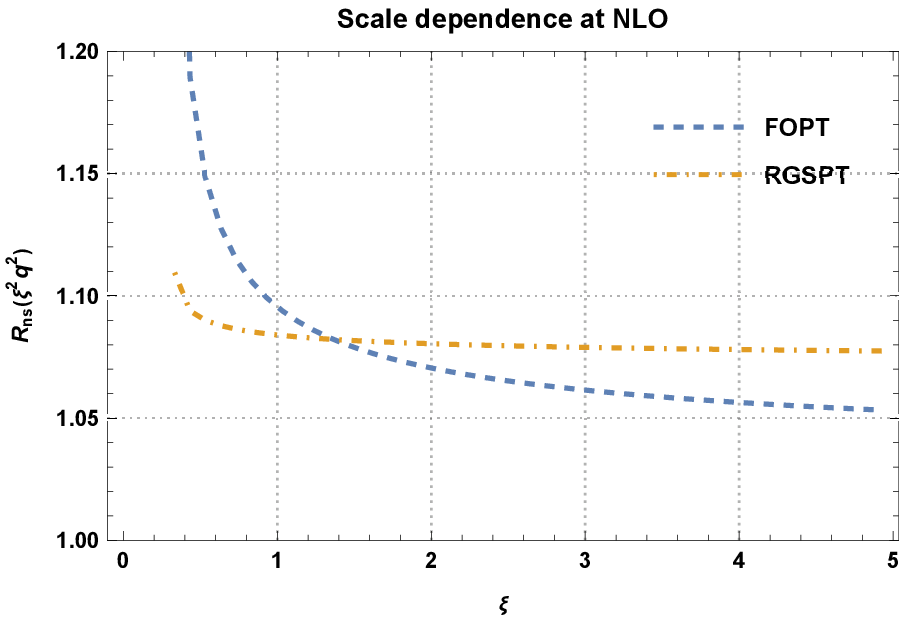}
				\includegraphics[width=.49\textwidth]{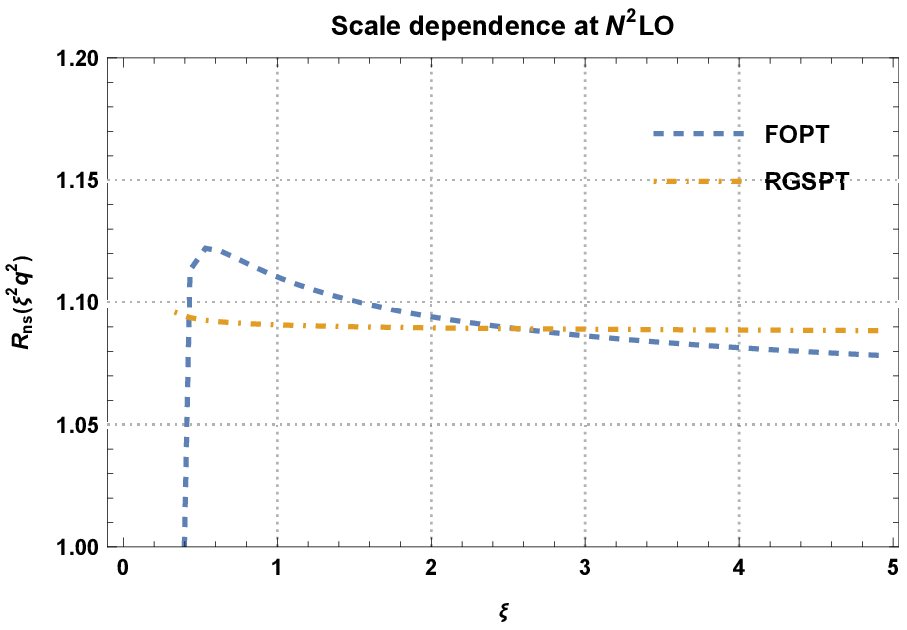}
				\includegraphics[width=.49\textwidth]{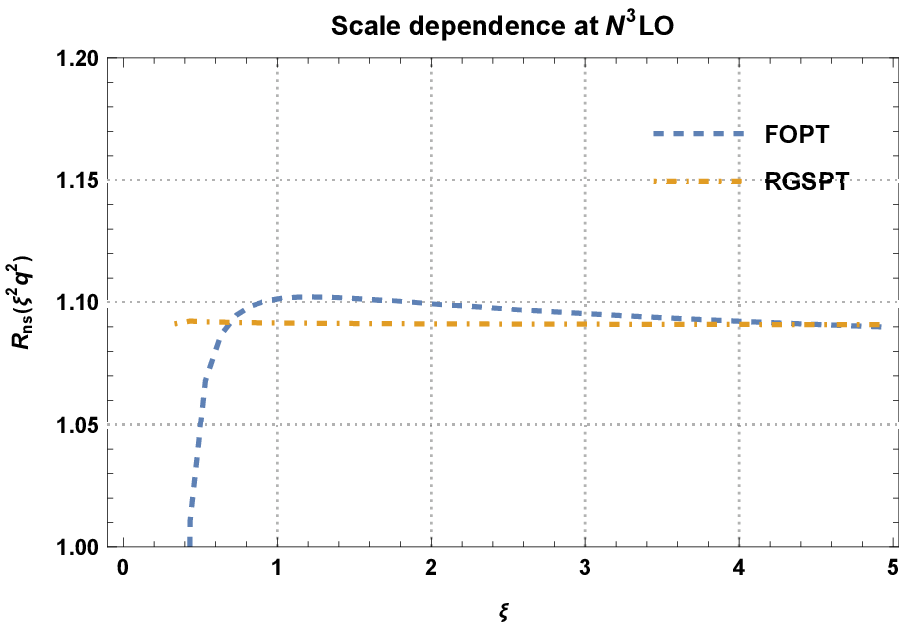}
				\includegraphics[width=.49\textwidth]{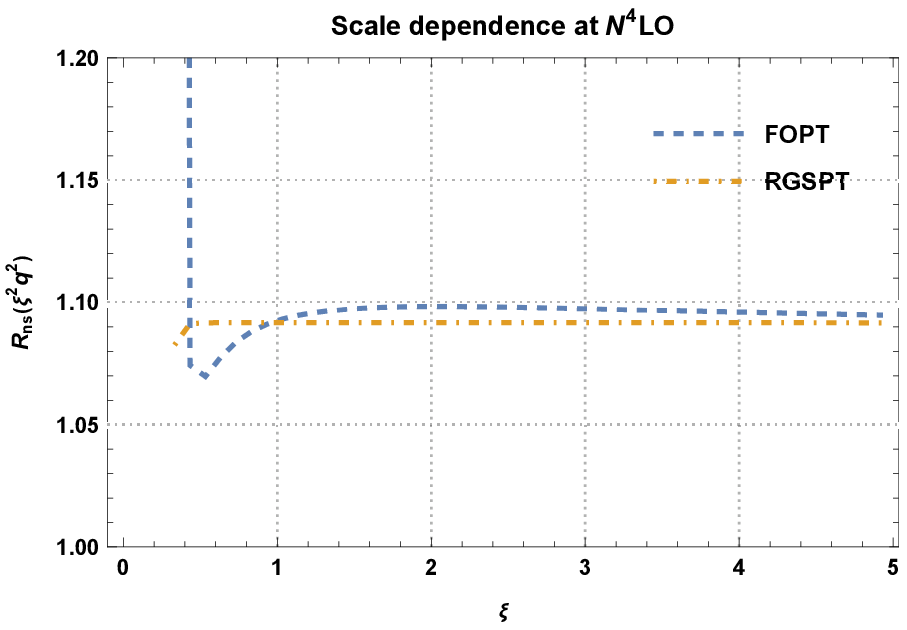}
			\caption{\label{fig:Rem2_scdep} The scale variation of $R_{em}(\xi^2 q^2)$ in the range  $\xi\in\left[1/3,5\right]$ for $n_f=3$  and $\alpha_s(q^2)=0.3$. }
		\end{figure}

 In this case, the leading massless Adler function in Eq.~\eqref{eq:adlerEM} receives contributions only from the non-singlet part of  Eq.~\eqref{eq:adler0}. Using $n_f=3$, scale variation $\xi\in[0.7,5]$  and $\alpha_s(q^2)=0.3$, the FOPT result for $R_{em}(s)$ has following contributions from various orders:
 \begin{align}
 	R_{em}^{\text{(I)}}(s)&=1.0+0.3183 \alpha _s+0.1661 \alpha _s^2-0.3317 \alpha _s^3-1.0972 \alpha _s^4+\order{\alpha_s^5}\nonumber\\&=1.0+ 0.0955+ 0.0150 -0.0090 -0.0089+\cdots\nonumber\\&=1.0926\pm0.0089_{\text{trunc.}}\pm0.0106_{\mu}=1.0926\pm0.0138\,,
 \end{align}
and the corresponding result for RGSPT reads:
\begin{align}
	R_{em}^{\text{(I)},\Sigma}(s)&=1+0.279995 \alpha _s+0.076018 \alpha _s^2+0.0285125 \alpha _s^3+0.022963 \alpha _s^4+\order{\alpha_s^5}\nonumber\\&=1+ 0.0840+ 0.0068+0.0008+ 0.0002+ \cdots \nonumber\\&=1.0918\pm0.0002_{\text{trunc.}}\pm0.00005_{\mu}=1.0918\pm0.0002\,.
\end{align}
	The total uncertainty in the RGSPT is $\sim$ 47 times smaller than the FOPT result because of the summation of the kinematical $\pi^2$-terms. The scale dependence in the two schemes can be seen in Fig.~\eqref{fig:Rem2_scdep}. \par
\begin{figure}[H]
              \centering
            \includegraphics[width=.7\textwidth]{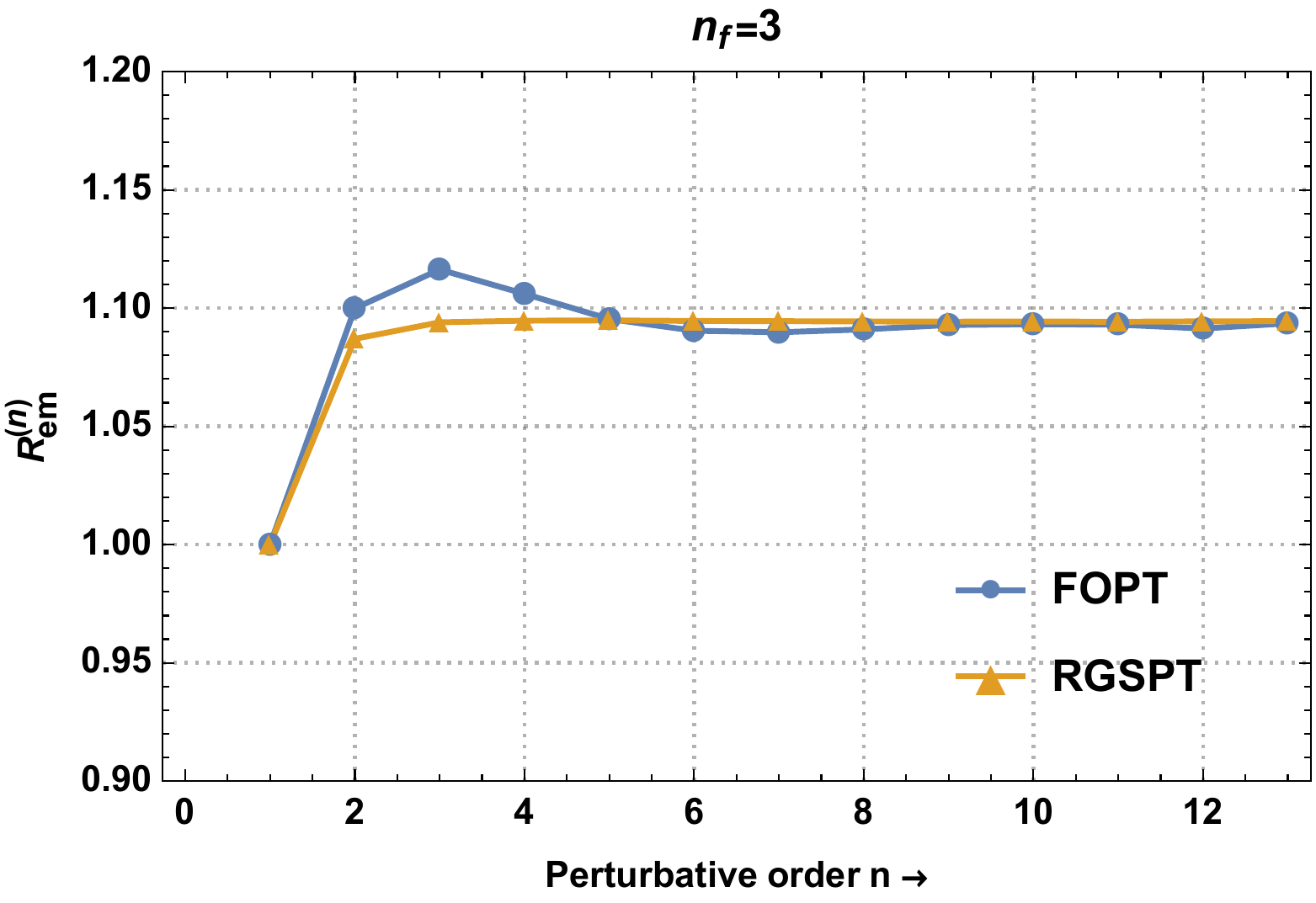}
           \caption{\label{fig:Rem}Stability of the $R_{em}$ for $n_f=3$ in the RGSPT and FOPT schemes.}
    \end{figure}

 
 We can also study the higher-order behavior using the results of Ref.~\cite{Beneke:2008ad} where the predictions for the higher coefficients in the case of hadronic $\tau$ decays using the Borel sum are obtained. These terms are collected in Eq.~\eqref{eq:AdlerTau} in the appendix~\eqref{app:Rem}. The higher-order corrections have the following numerical values:
 \begin{align}
     R_{em}^{\text{(I)}}=&1+0.09549+0.01495-0.00896 -0.00889-0.00394 \nonumber\\&\hspace{2mm}-0.00050+ 0.00062+ 0.00113+ 0.00052+ 0.00025 -0.00103+ 0.00095\,,\nonumber\\
     R_{em}^{\text{(I)},\Sigma}=&1+0.08400+ 0.00684+ 0.00077+ 0.00019 -0.00019\nonumber\\&\hspace{2mm}-0.00007 -0.00009-0.00009+ 0.00004-0.00010+ 0.00015-8\times10^{-6}\,.
 \end{align}
 Interestingly, the kinematical terms dominate the genuine contributions from N$^5$LO for RGSPT. These effects are relatively smaller compared to the FOPT, and reliable predictions from a truncated series using RGSPT can still be obtained. As shown in Fig.~\eqref{fig:Rem}, higher order contributions are under control. 

\subsubsection{Case-II: Four Active Flavors}\label{subsub:rem4}
For this case, $n_f=4$, and we use only the non-singlet contributions known to $\ordas{4}$. Using $\alpha_s(q^2)=0.2$, the R-ratio has the following numerical form:
\begin{align}
	R_{em}^{II,\text{FOPT}}(s)=&1+0.3183 \alpha _s+0.1545\alpha _s^2-0.3715\alpha _s^3-0.9536 \alpha _s^4+\order{\alpha_s^5}\nonumber\\=&1.0+ 0.0637+ 0.0062 -0.0030 -0.0015+\cdots\nonumber\\=&1.0653\pm0.0015_{trunc.}\pm0.0010_{\mu}=1.0653\pm0.0018\,,
	\label{eq:Rem_fixed_1}
\end{align}
and the kinematical $\pi^2-$terms starts dominating from $\ordas{3}$. The uncertainty from the scale variations is obtained by varying $\xi$ in the range~$\xi\in\left[1/2,5\right]$.\par
\begin{figure}[ht]
  \centering
 				\includegraphics[width=.49\textwidth]{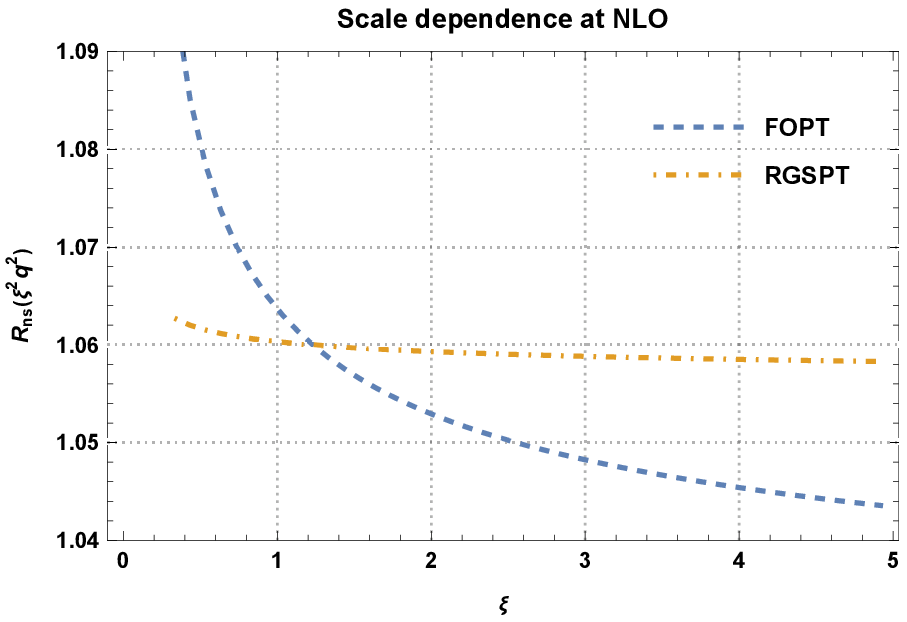}
 			\includegraphics[width=.49\textwidth]{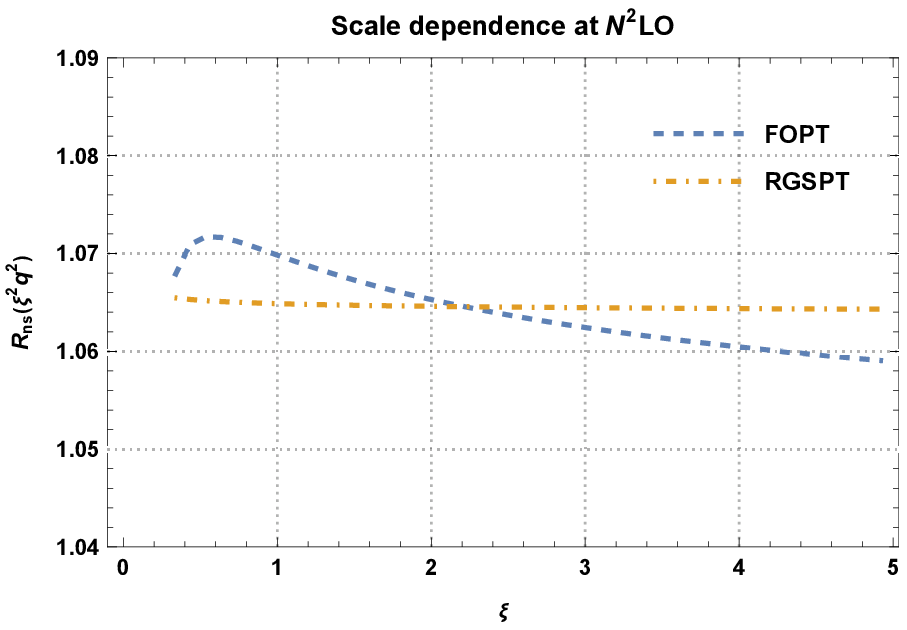}
 			\includegraphics[width=.49\textwidth]{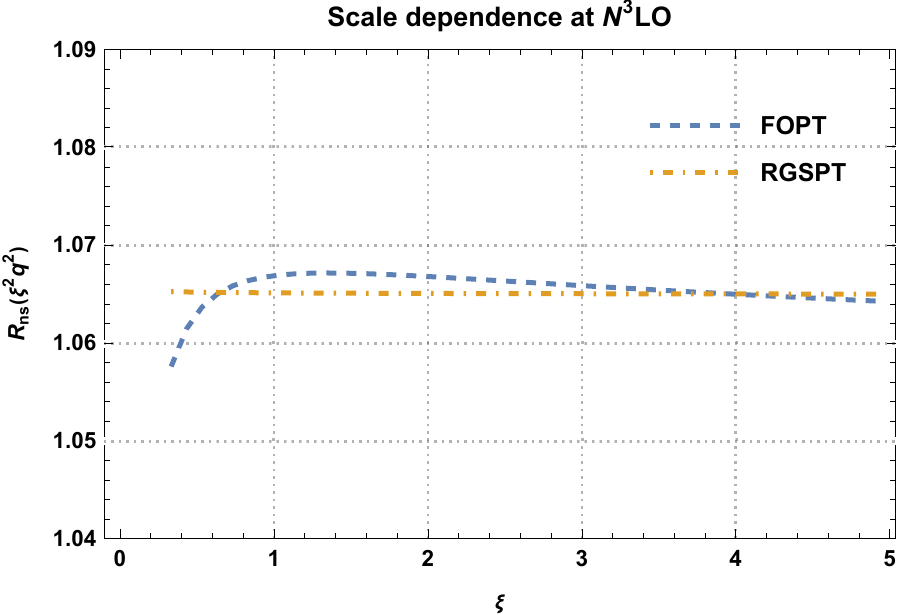}
 			\includegraphics[width=.49\textwidth]{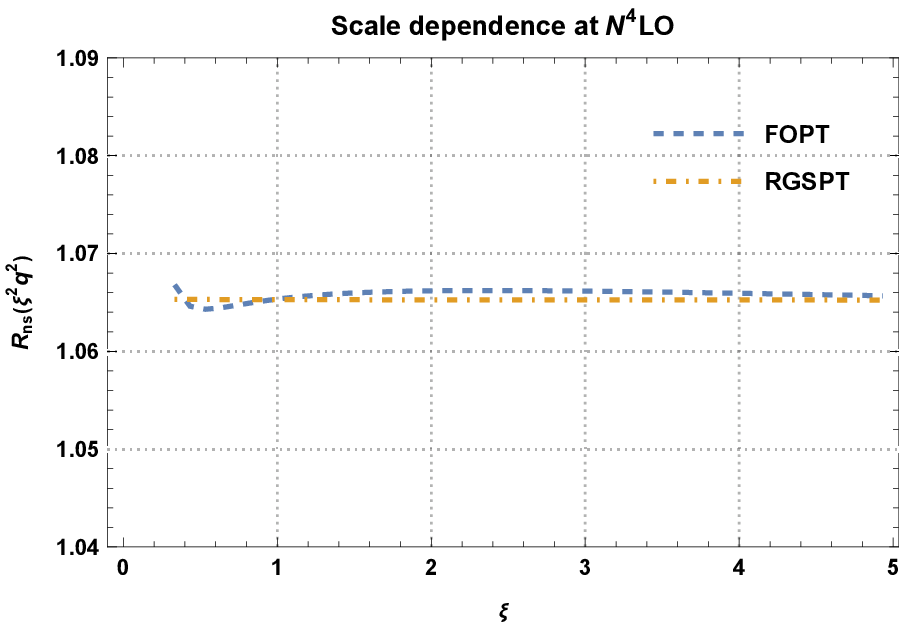}
 		\caption{\label{fig:Rem1_scdep} The scale variation of the non-singlet part of the $R_{em}(\xi^2 q^2)$ in the FOPT and RGSPT schemes when scale parameter $\xi$ is varied in the range  $\xi\in\left[1/3,5\right]$ for $n_f=4$ using $\alpha_s(q^2)=0.2$.}
 	\end{figure}

Using the same inputs as in Eq.~\eqref{eq:Rem_fixed_1}, the electromagnetic R-ratio in the RGSPT scheme has the following numerical form:
\begin{align}
	R_{em}^{\left(II\right),\text{RGSPT}}(s)&=1+0.3016 \alpha _s+0.1140 \alpha _s^2+0.0293 \alpha _s^3+0.1086 \alpha _s^4+\order{\alpha_s^5}\nonumber\\&=1.0+ 0.0603+ 0.0046+ 0.00023+0.00017+\cdots\nonumber\\&=1.0653\pm0.0002_{\text{trunc.}}\pm0.00005_{\mu}=1.0653\pm0.0002,.
	\label{eq:Rem_summed_1}
\end{align}
 We can see from Eq.~\eqref{eq:Rem_fixed_1} and Eq.~\eqref{eq:Rem_summed_1} that the resummation of the kinematical term improves the convergence as well as the scale dependence for RGSPT compared to the case when FOPT is used. The theoretical uncertainty in the RGSPT results in Eq.~\eqref{eq:Rem_summed_1}) is $\sim$1/10 smaller than that of FOPT result in Eq.~\eqref{eq:Rem_fixed_1}. The scale dependence is minimal after including the N$^4$LO result in FOPT. However, the RGSPT always has better control over the scale dependence than FOPT at each order which can be seen in Figs.~\eqref{fig:Rem1_scdep}.

\begin{figure}[H]
              \centering
            \includegraphics[width=.7\textwidth]{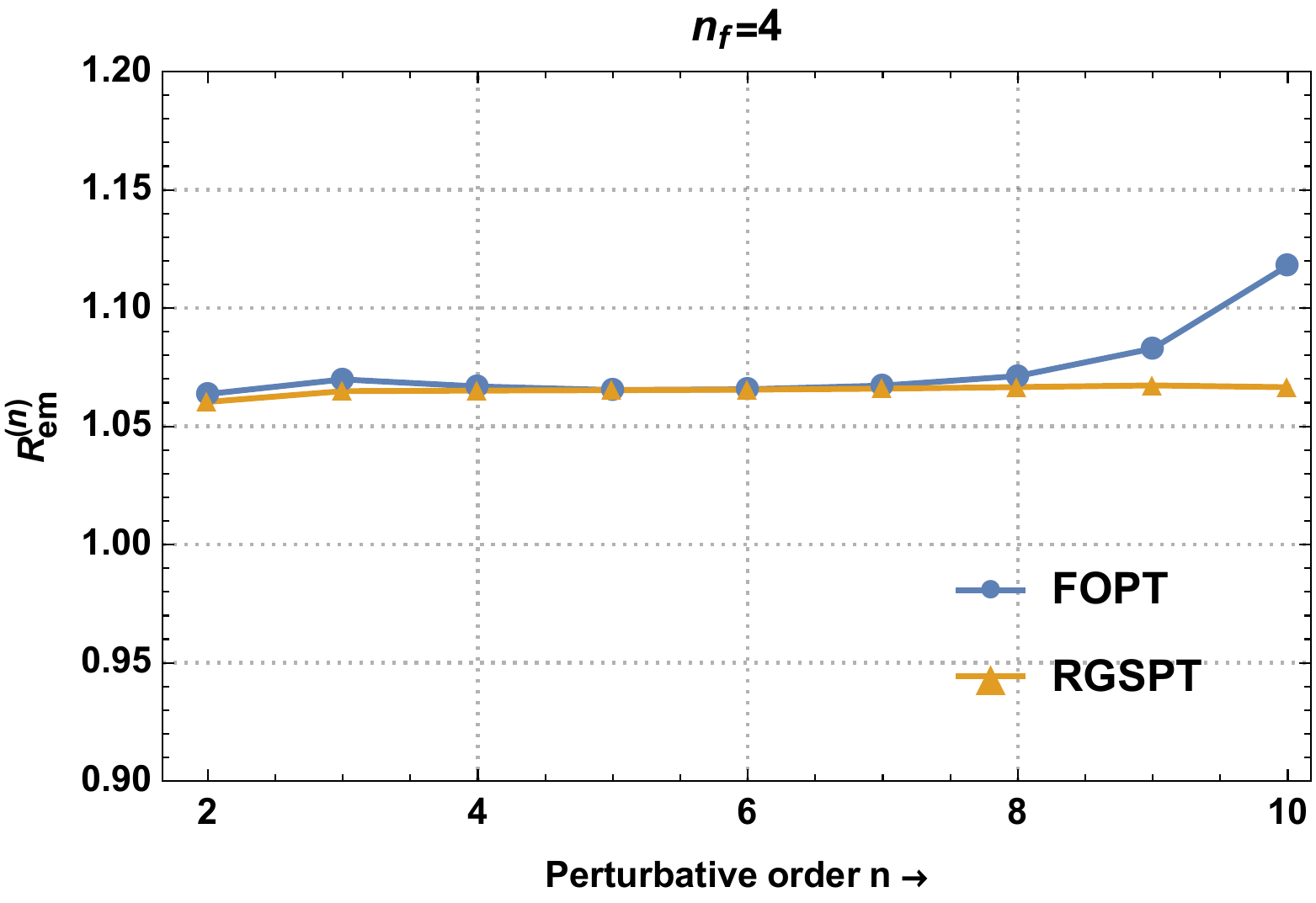}
           \caption{\label{fig:Rem4} Higher order behavior of the non-singlet contribution to $R_{em}$ for $n_f=4$ using $\as=0.2$.}
    \end{figure}

To the best of our knowledge, there are no predictions for the higher-order coefficients in this case. We use the Pad\'e approximants as discussed in the section~\eqref{subsec:hgg}. Using the $\ordas{4}$ coefficients of non-singlet Adler function from Eq.~\eqref{eq:adler0}, following Pad\'e approximants can be constructed:
\begin{align}
   W^{\left(0,4\right)} =&\frac{1}{1 -  x - 0.524526 x^2 - 0.709564 x^3 - 
 23.121 x^4},\\W^{\left(1,3\right)} =& \frac{1 - 32.5847 x}{1 - 33.5847 x + 
 32.0602 x^2 + 16.382 x^3}\,,\\ W^{\left(2,2\right)} =&\frac{1 - 52.3638 x + 
 26.7566 x^2}{1 - 53.3638 x + 78.5959 x^2}\,,\\ W^{\left(3,1\right)} =&\frac{1 - 
 8.92846 x - 8.40393 x^2 - 12.3776 x^3}{1 - 9.92846 x}\,,
\end{align}
leading to the following average:
\begin{align}
    \overline{W}=&1 +  x + 1.52453 x^2 + 2.75862 x^3 + 27.3888 x^4+ 594.038 x^5 + 
 23309.3 x^6 \nonumber\\&\hspace{2mm}+ 1.05282*10^6 x^7  + 5.01986\times10^7 x^8 + 
 2.46615\times10^9 x^9 + 1.23479\times10^{11} x^{10}\,.
\end{align}
 Using these coefficients, the higher-order behavior of $R_{em}$ using $\as=0.2$ for FOPT are obtained as:  
\begin{align}
    R_{em}^{\text{FOPT}}=&1+ x+ 1.52453 x^2-11.5203 x^3 -92.891 x^4+ 308.683 x^5+ 23278.8 x^6\nonumber\\&\hspace{2mm}+ 954842 
    x^7+4.31945\times10^7 x^8+ 2.03702\times10^9 x^9+9.73859\times10^{10} x^{10}+\order{x^{11}}\nonumber\\=&1+ 0.06366+ 0.00618-0.00297 -0.00153+ 0.00032+ 0.00155\nonumber\\&\hspace{2mm}+ 0.00405+0.01165+ 0.03499+ 0.10649+\cdots\,,
\end{align}
and for RGSPT, we obtain the following contributions:
\begin{align}
     R_{em}^{\text{RGSPT}}=&1+0.947499 x+ 1.12508 x^2+ 0.909611 x^3+ 10.5751 x^4+ 221.145 x^5+6241.56 x^6\nonumber\\&\hspace{2mm}+ 157010 x^7+2.61906\times10^6 x^8 -4.44785\times10^7 x^9 -7.62188\times10^9 x^{10}\nonumber\\=&1+0.06032+ 0.00456+0.00023+ 0.00017+ 0.00023+ 0.00042\nonumber\\&\hspace{2mm}+0.00067+0.00071 -0.00076 -0.00833+\cdots\,.
\end{align}
The higher-order behavior using the above results for FOPT and RGSPT are shown in Fig.~\eqref{fig:Rem4}. It is evident that the RGSPT results for $R_{em}$ are more stable than FOPT if we include the higher order predictions using the Pad\'e approximants. In FOPT, these higher-order contributions are significantly larger than RGSPT.  

\subsubsection{Light Quark Hadronic Contribution Muon \texorpdfstring{$g-2$}{}}
Muon anomalous magnetic moment anomaly has been of constant interest in recent years. The tension between the predictions of the standard model and experiments now stands at $4.2\sigma$ ~\cite{Muong-2:2021ojo,Muong-2:2006rrc,Aoyama:2020ynm}. The key issue of the anomaly is contributions coming from the hadron vacuum polarization (HVP). Lattice QCD and dispersive approach based on data-driven methods are used in the literature to quantify these HVP contributions. More details on the subject can be found in the Refs.~\cite{Colangelo:2022jxc,Boito:2022dry,Golterman:2022jvz,Boito:2022njs,Ananthanarayan:2018nyx,Ananthanarayan:2022wsl,Ananthanarayan:2023gzw} and references therein. \par %
 In this subsection, we will focus only on the pQCD contribution to HVP relevant to the summation of the kinematical $\pi^2-$terms. Such contributions enter while evaluating the leading order continuum contributions to HVP using the pQCD input. The electromagnetic R-ratio discussed in section~\eqref{subsub:rem3} with some massive corrections are taken as inputs to calculate these contributions using the following relation:
\begin{align}
	a_\mu^{\text{LO},\text{HVP}}=&\frac{1}{3}\frac{\alpha^2_{em}}{\pi^2}\int_{s_0}^{\infty}\frac{G_2(s)}{s}R_{em}(s)\,,
\end{align}
where kernel $G_2(s)$ is given by ~\cite{Nesterenko:2021byp,Aoyama:2020ynm}:
\begin{align}
G_2	(s)=\int_{0}^{1}dz\frac{(z^2 (1 - z))}{(z^2 + (1 - z) s/m_\mu^2)}\,,
\end{align}
and $m_\mu= 105.6583745\MeV$ is the mass of Muon. Recently, three-flavor s-quark connected and disconnected contributions to Muon $g-2$ are calculated in Ref.~\cite{Boito:2022rkw} using the FOPT scheme. These contributions are denoted as $\tilde{a}_{\mu}^{\text{cont.}}$ in this article. \par
\begin{figure}[H]
\centering
\begin{subfigure}{.49\textwidth}
	\includegraphics[width=\textwidth]{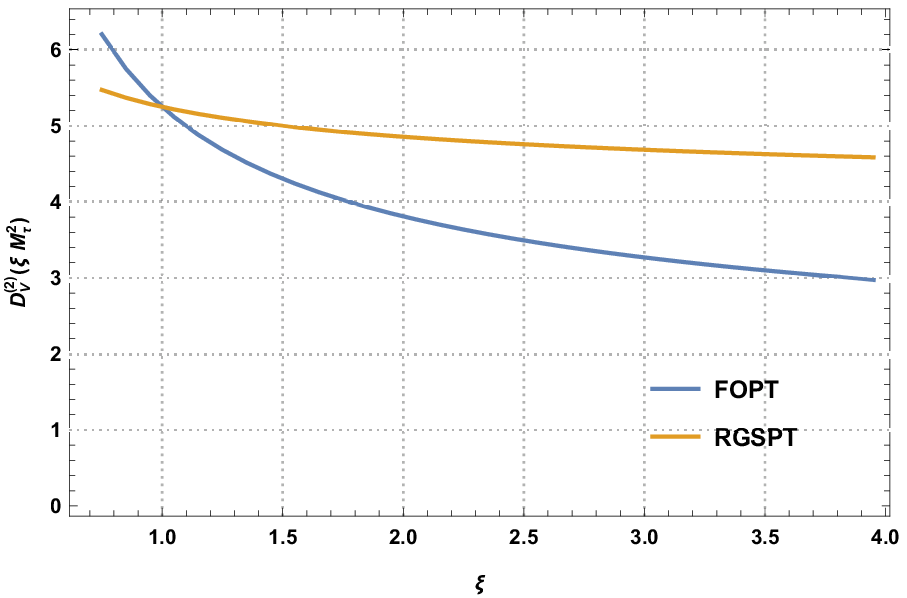}
 \caption{\text{ \label{fig:D0V}}}
 \end{subfigure}
 \begin{subfigure}{.49\textwidth}
     \includegraphics[width=\textwidth]{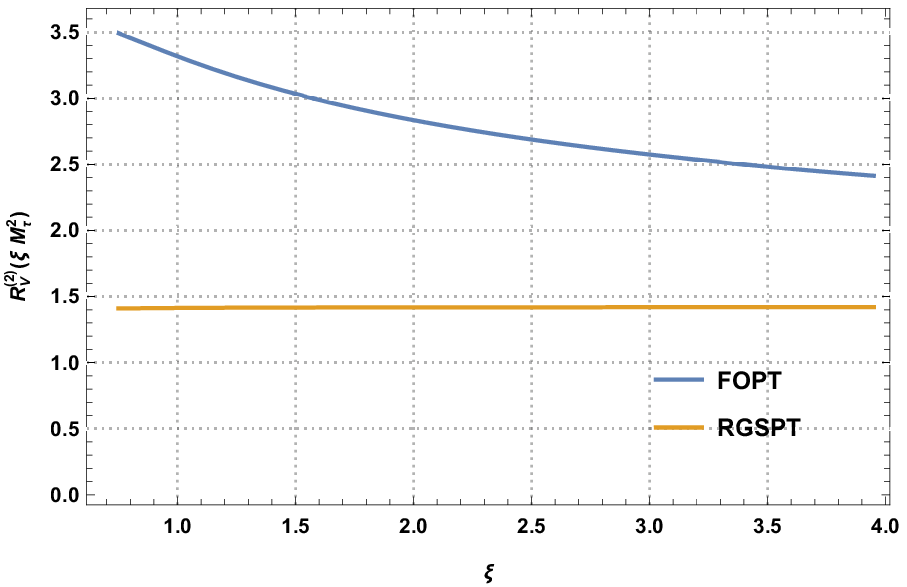}
    \caption{\label{fig:R0V}}
 \end{subfigure}
 \begin{subfigure}{.49\textwidth}
     \includegraphics[width=\textwidth]{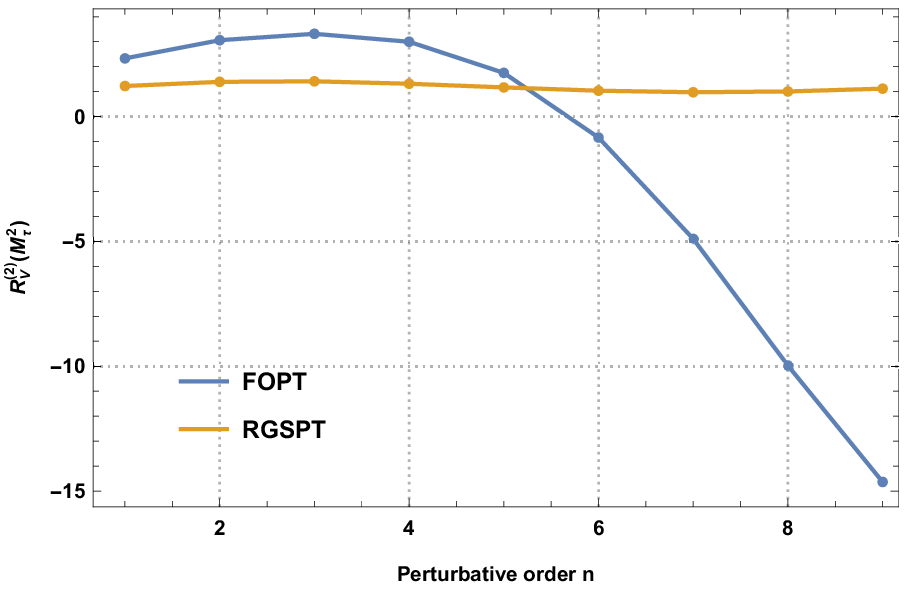}
     \caption{\label{fig:R0nV}}
 \end{subfigure}
 \caption{The scale dependence of the $D_V$ and $R_{em}$ using $\ordas{4}$ corrections in (i) %\eqref{fig:D0V}%
 and (ii), %\eqref{fig:R0V}
 respectively. The stability of the $R_{em}$ for higher-order as using Pad\'e predictions in (iii)%\eqref{fig:R0nV}
\label{fig:DRV} }
\end{figure}

 The effects of the summation of $\pi^2-$terms to the $R_{em}$ from massless Adler function and higher order behavior are already discussed in the section~\eqref{subsub:rem3}. However, some small corrections are also due to finite strange quark mass ($m_s$). These massive corrections for the Adler function are already known to $\ordas{4}$ 
 from Refs.~\cite{Chetyrkin:1990kr,Chetyrkin:1994ex,Chetyrkin:2000zk,Baikov:2004ku} and can be found in appendix~\eqref{app:Rem_mass_correction}. Higher-order corrections can also be predicted using the Pad\'e approximants. Following the procedure opted in the previous subsection, the Adler function in the massive case is obtained as:
\begin{align}
D_{V,2}^{\text{Pad\'e}}=&12 x + 113.5 x^2 + 1275.89 x^3 + 16496.5 x^4 + 215732 x^5 + 2.87888\times10^6 x^6\nonumber\\&\hspace{.63cm} + 3.91033\times10^7 x^7 + 5.3773\times10^8 x^8 +7.47708\times 10^9 x^9+\cdots\,.
\end{align} 
Now, we can use the above coefficients to study the stability of the $R_{em}^{\left(2\right)}$ with respect to the renormalization scale and with the order of the perturbation theory.  The $R_{em}^{\left(2\right)}$ for FOPT, using $\as(\mtsq)=0.3139$, is obtained as:
\begin{align}
   R_{em}^{\left(2\right)}=& 12 x + 113.5 x^2 + 730.597 x^3 + 2569.98 x^4 - 31969.7 x^5 - 
 1.25\times10^6 x^6 \nonumber\\&- 2.61\times10^7 x^7 - 4.08729\times10^8 x^8 - 
 5.1128\times10^9 x^9 - 4.7013\times10^{10} x^{10}\nonumber\\& - 1.32053\times10^{11} x^{11}+\order{x^{12}}\\
 =&1.199+ 1.1331+ 0.7288+ 0.2562 -0.3184 -1.2469-2.5910 -4.0604\nonumber\\&-5.0750 -4.6626 -1.3086+\cdots\,.
\end{align}
For the RGSPT scheme, 
\begin{align}
R_{em}^{\left(2\right),\Sigma}=&8.2999 x+ 39.6882 x^2+ 167.085 x^3+ 214.167 x^4 -9709.35 x^5
-148249. x^6 \nonumber\\&-1.32872\times10^6 x^7 -6.2473\times10^6 x^8+ 3.26504\times10^7 x^9+ 
1.13115\times10^9 x^{10}\nonumber\\&+ 1.44578\times10^{10} x^{11} +\order{x^{12}}\,,\nonumber\\=& 0.8293+ 0.3962+ 0.1667+ 0.0213 -0.0967 -0.1475 -0.1321\nonumber\\& -0.0621+
0.0324+ 0.1122+ 0.1433+\cdots\,.
\end{align}
From these numerical predictions, it is clear that RGSPT has better convergence than FOPT for the known $\ordas{4}$ results but does not have a convergent behavior when higher order coefficients are used. For FOPT, it is a much more serious issue than RGSPT, as shown in Fig.~\eqref{fig:R0nV}.
The scale dependence for massive corrections to the Adler function and $R_{em}(s)$ in the timelike region be found in Fig.~\eqref{fig:DRV}. These quantities are calculated using $\alpha_s(\mtsq)=0.3139$ for three flavors and used in the rest of the subsection. It is clear from these plots that the pQCD analysis for the low energy $s\sim\mtsq$ requires summation in order to make meaningful predictions. The RGSPT series shows that the summation of kinematical $\pi^2-$terms improves the scale dependence and the radius of convergence for known $\ordas{4}$ terms. 
\par
  \par
\begin{figure}[H]
\centering
			\begin{subfigure}{.48\textwidth}
			\includegraphics[width=\textwidth]{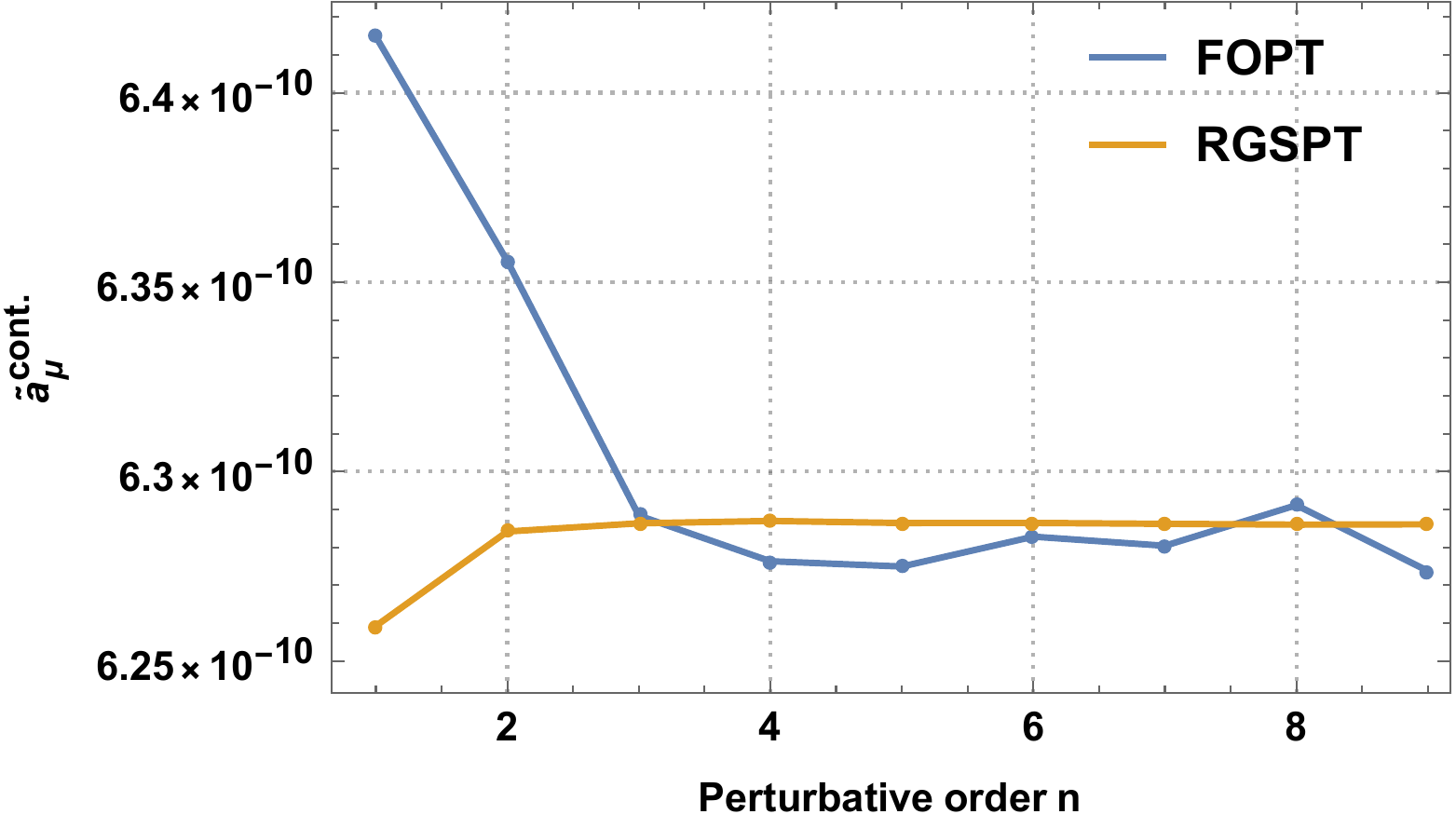}
  \caption{ \label{fig:muon}}
			\end{subfigure}
			\begin{subfigure}{.48\textwidth}
				\includegraphics[width=\textwidth]{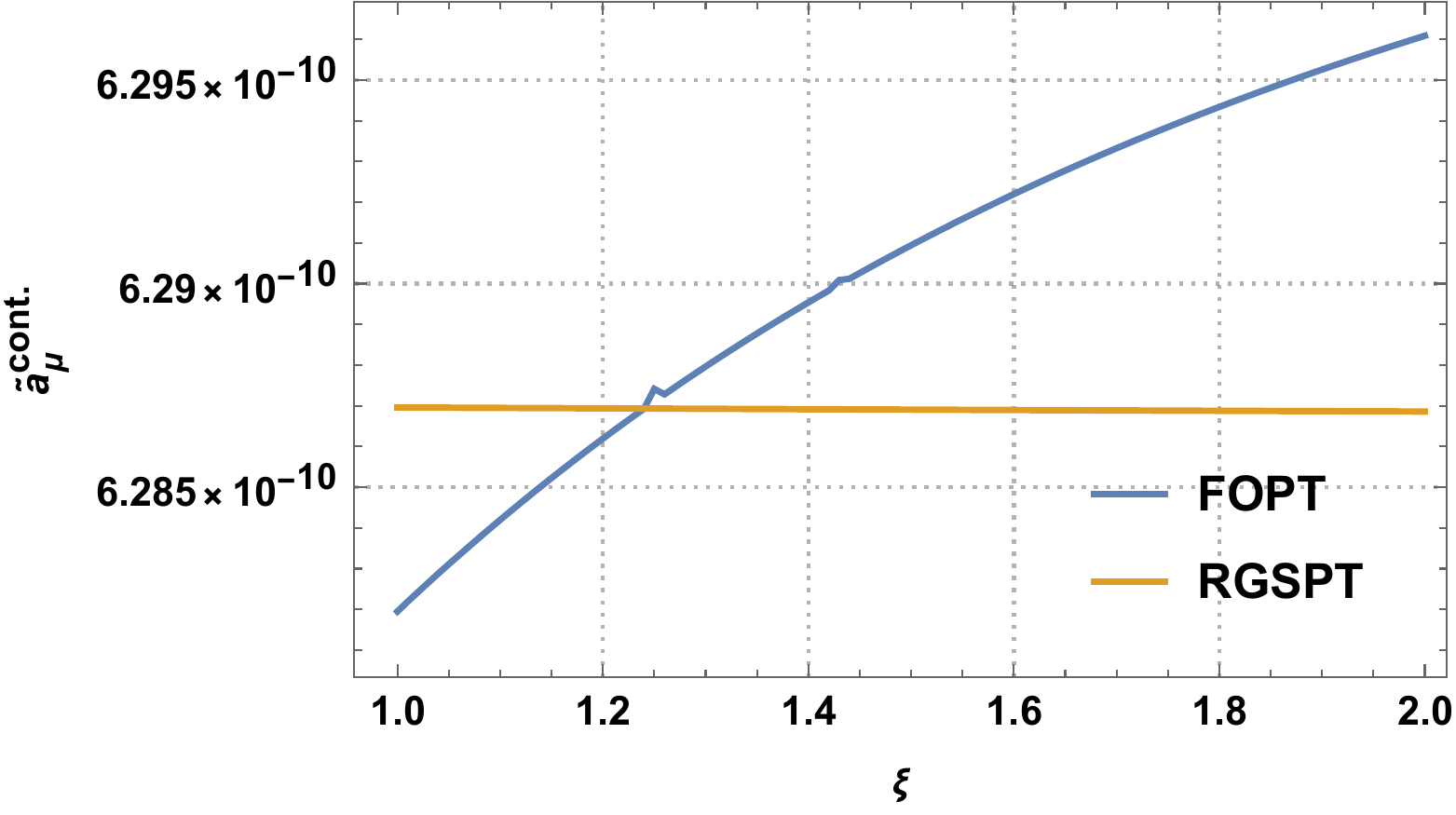}
   \caption{ \label{fig:muon0_scdep} }
			\end{subfigure}
   \begin{subfigure}{.48\textwidth}
				\includegraphics[width=\textwidth]{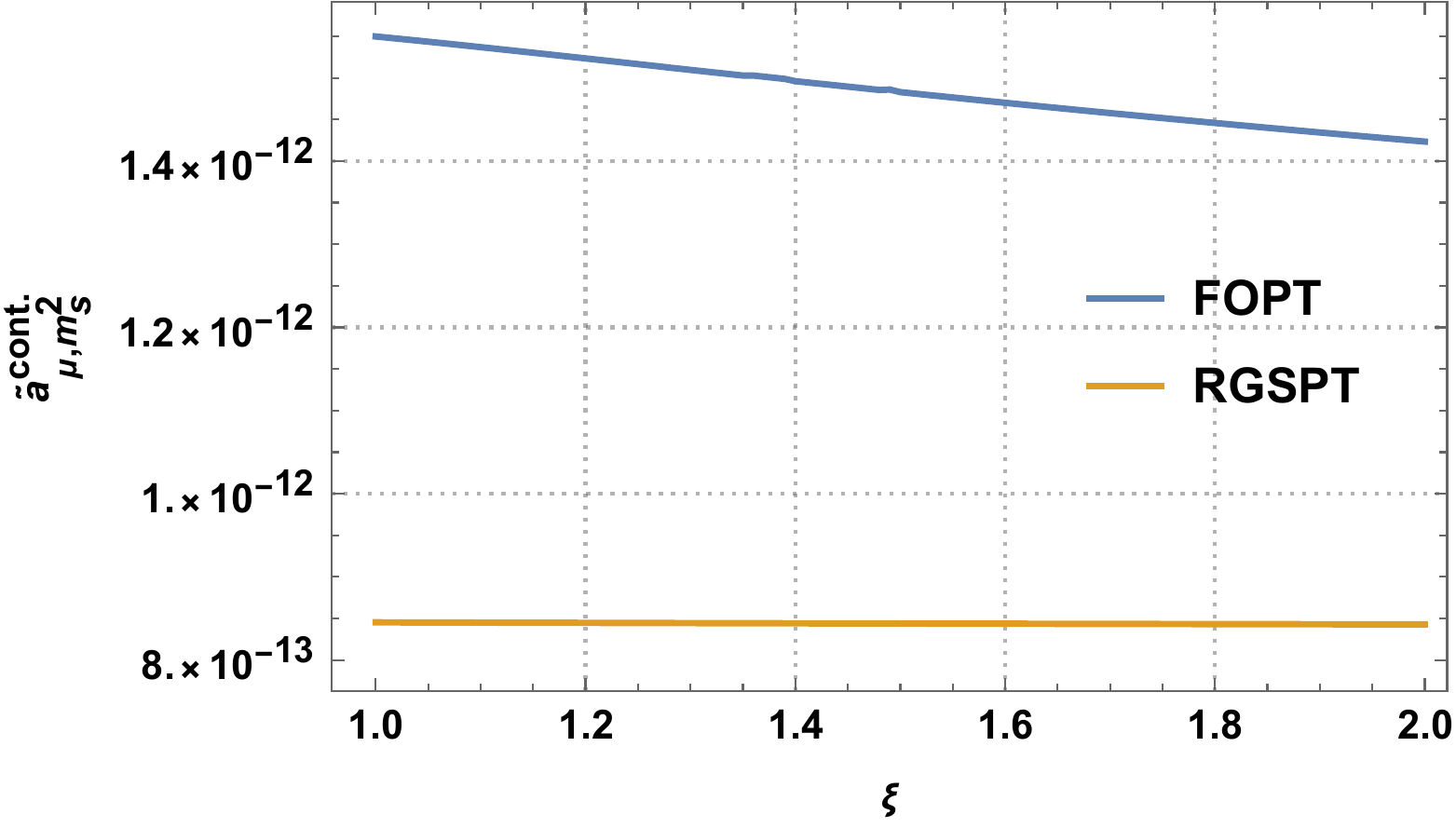}
   \caption{ \label{fig:muon2_scdep} }
			\end{subfigure}
   \caption{The behavior of $\tilde{a}_{\mu}^{\text{cont.}}$ with increasing perturbative order in $\ordas{4}$ in Fig.~\eqref{fig:muon} and scale dependence for the exactly known terms in Fig.~\eqref{fig:muon0_scdep}. The scale dependence of the massive corrections to $\tilde{a}_{\mu,m_s^2}^{\text{cont.}}$ in Fig.~\eqref{fig:muon2_scdep}. }
   \end{figure}

 
  The calculation of $\tilde{a}_{\mu}^{\text{cont.}}$ requires both massless and massive strange quark mass correction, and relevant expressions are collected in appendix~\eqref{app:Rem}.
  Before moving to $\tilde{a}_{\mu}^{\text{cont.}}$ calculation, it should be noted $\order{\alpha_s^5}$ coefficient to massless correction, $d_5=283$,  provided in Ref.~\cite{Beneke:2008ad} is used to estimate the truncation uncertainties. Using FOPT, we obtain the following contribution to $\tilde{a}_{\mu}^{\text{cont.}}$:
\begin{align}
		\tilde{a}_{\mu}^{\text{cont.}}=\begin{cases}
			&6.277\times10^{-10} \text{ including $\order{\alpha^5_s}$}\\
			&6.282\times 10^{-10} \text{ without $\order{\alpha^5_s}$}
		\end{cases}\,,
\end{align}
 and for RGSPT, we get the following contributions:
 \begin{align}
 	\tilde{a}_{\mu}^{\text{cont.}}=\begin{cases}
 		&6.286\times10^{-10} \text{ including $\order{\alpha^5_s}$}\\
 		&6.287\times10^{-10} \text{ without $\order{\alpha^5_s}$}
 	\end{cases}\,.
 \end{align}
We can see that the truncation error is significantly reduced in the RGSPT. The scale dependence of $\tilde{a}_{\mu}^{\text{cont.}}$ using known $\ordas{4}$ is plotted in Fig.~\eqref{fig:muon0_scdep}. Higher-order terms obtained in Beneke and Jamin~\cite{Beneke:2008ad} can be used to study the behavior of $\tilde{a}_{\mu}^{\text{cont.}}$ in different schemes. The numerical values of these coefficients are presented in the appendix~\eqref{app:Rem}. The numerical stability of the $\tilde{a}_{\mu}^{\text{cont.}}$ using these coefficient is presented in Fig.~\eqref{fig:muon}.  

We can also include the massive corrections from the strange quark mass represented as $\tilde{a}_{\mu,m_s^2}^{\text{cont.}}$. The expressions for such corrections in the FOPT and RGSPT schemes are presented in the appendix~\eqref{app:Rem_mass_correction}. Using the strange quark mass as $m_s(2\GeV)=93.4\pm8.6\MeV$ as input from the PDG\cite{Workman:2022ynf}, we obtain the following corrections to $\tilde{a}_{\mu,m_s^2}^{\text{cont.}}$: 
\begin{align}\tilde{a}_{\mu,m_s^2}^{\text{cont.}}&=1.550\times10^{-12}\quad\text{(FOPT)}\,,\\
\tilde{a}_{\mu,m_s^2}^{\text{cont.}}&=8.453\times 10^{-13}\quad\text{(RGSPT)}\,.
\end{align}
The RGSPT contribution is~$\sim54.5\%$ smaller than the FOPT due to the enhanced suppression by the presence of the quark mass anomalous dimension in the denominator. Even though massive corrections to $\tilde{a}_{\mu}^{\text{cont.}}$ are small, analytic continuation effects are numerically significant in this case. The scale dependence of these contributions is presented in Fig.~\eqref{fig:muon2_scdep}.
\section{Summary and Conclusion}\label{sec:summary}
Effects of kinematical corrections arising from analytic continuation in QCD have a long history. There are various methods used in the literature to control their effects. These effects are related to the RG behavior of the perturbative series. RGSPT uses RGE to sum running logarithms originating from a given order of perturbation series to all orders. We have used this property to sum these kinematical terms to all orders in addition to the reduced dependence on the renormalization scale. These issues are discussed in section~\eqref{sec:intro_RGSPT} and section~\eqref{sec:def_RD} leading to important all order summation formula in RGSPT in Eq.~\eqref{eq:master_eq}.\par
In section~\eqref{sec:applications}, the effects of the kinematical terms using FOPT and RGSPT for various processes involving Higgs decay~\eqref{subsec:Hdecay} in pQCD and electromagnetic R-ratio for $e^+e^-$~\eqref{subsec:Rem} are discussed.\par 
In section~\eqref{subsec:Hbb}, Higgs decaying to a pair of bottom quarks is studied. The results obtained from RGSPT show enhanced stability with respect to scale variation in addition to good convergence, even if predictions for higher orders are included. For FOPT, cancellation between genuine perturbative correction and $\pi^2$-terms occurs, leading to less truncation uncertainty for the known $\ordas{4}$ results. The truncation uncertainty for the higher order term should be estimated from the Adler function rather than analytically continued series in FOPT if kinematical $ \pi^2 $ -terms are involved. \par In section~\eqref{subsec:hgg}, we study the decay of Higgs to a pair of gluons. The analytically continued series has a better convergence to known orders using RGSPT than FOPT. However, in the total decay width, contributions from the Wilson coefficients result in slightly greater truncation uncertainty for RGSPT. \par
In section~\eqref{subsec:H_Hadron}, the results from the previous two subsections as well as some contributions from the mixing of gluonic and fermionic operators, are also included. We get to observe the same patterns as discussed in the previous two subsections.\par 
In section~\eqref{subsec:Rem}, electromagnetic R-ratio for $n_f=3$ and $n_f=4$ are studied. Its application in continuum contribution to muon $g-2$ is discussed, including the massive correction from the s-quark. In these examples, a significant reduction in the truncation uncertainties is also achieved in addition to improved scale dependence. We believe that these results can be used to calculate the experimental values of the low-energy moments of the vector current correlators~\cite{Kuhn:2007vp,Dehnadi:2011gc,Dehnadi:2015fra,Boito:2019pqp,Boito:2020lyp} that are important observables in the determination of the $\as$, charm- and bottom- quark masses. We have used the experimental moments available in the determination of the $\as$, charm- and bottom-quark masses in Ref.~\cite{AlamKhan:2023kgs}.\par
In addition, the results obtained in this article can be used in the QCD sum rules where analytic continuation of the polarization function is required. One such application we have used in the light quark mass determination from the divergences of the axial vector current in Ref.~\cite{AlamKhan:2023ili}.

\section{Acknowledgments}
We thank Prof. B. Ananthanarayan and Mr. Aadarsh Singh for carefully reading the manuscript and their valuable comments. Author is also thankful to Prof. Apoorva Patel for the financial support. The author is also supported by a scholarship from the Ministry of Human Resource Development (MHRD), Govt. of India. This work is a part of the author's Ph.D. thesis.

	\appendix

   \section{Running of the Strong Coupling and the Quark Masses in the pQCD. }\label{app:mass_run}

	The evolution of the strong coupling and quark masses is computed by solving the following RGEs:
	\begin{align}
		\mu^2\frac{d}{d\mu^2}x(\mu)&=\beta(x)=-\sum_{i=0} (x(\mu))^{i+2} \beta_i\,,\nonumber \\
		\mu^2 \frac{d}{d\mu^2}m(\mu)&\equiv\hspace{2mm}m(\mu) \hspace{.4mm}\gamma_m(x(\mu) =	-m(\mu)\sum_{i=0}(x(\mu))^{i+1} \hspace{.4mm} \gamma_i\label{anomalous_dim}\,.
	\end{align}	
 where the coefficients of QCD beta functions~\cite{vanRitbergen:1997va,Gross:1973id,Caswell:1974gg, Jones:1974mm,Tarasov:1980au,Larin:1993tp,Czakon:2004bu,Luthe:2016ima,Baikov:2016tgj,Herzog:2017ohr} and quark mass anomalous dimension~\cite{Tarrach:1980up,Tarasov:1982plg,Larin:1993tq,Vermaseren:1997fq,Chetyrkin:1997dh,Baikov:2014qja,Luthe:2016xec,Baikov:2017ujl} can be found in Ref.~\cite{Ananthanarayan:2022ufx}.

\section{Adler function for the vector correlator and electromagnetic current.}
\label{app:Rem}
The Adler function for the vector current correlators is known to  $\order{\alpha_s^4}$ in Ref.~\cite{Appelquist:1973uz,Zee:1973sr,Chetyrkin:1979bj,Dine:1979qh,Gorishnii:1990vf,Surguladze:1990tg,Chetyrkin:1996ez,Baikov:2008jh,Baikov:2010je,Herzog:2017dtz}. The Adler function for $e^+e^-$ also receives additional singlet contributions. The numerical expression for $n_f$ flavor is given by:
\begin{align}
   D_{\text{NS}} =&1 + x +x^2 [1.98571 + L (2.75 - 0.166667  n_f) -   0.115295  n_f] + x^3 [18.2427 - 4.21585  n_f \nonumber\\&+ 0.0862069  n_f^2 +  L^2 (7.5625 - 0.916667  n_f + 0.0277778  n_f^2) +
   L (17.2964 - 2.08769  n_f\nonumber\\&+ 0.0384318  n_f^2)] +x^4 [135.792 -
   34.4402  n_f + 1.87525  n_f^2 - 0.0100928  n_f^3 +
   L (198.14\nonumber\\& - 52.8851  n_f + 3.09572  n_f^2 - 0.0431034  n_f^3) +
   L^2 (88.8789 - 16.1754  n_f + 0.812399  n_f^2 \nonumber\\&- 0.00960795  n_f^3) +
   L^3 (20.7969 - 3.78125  n_f + 0.229167  n_f^2 - 0.00462963  n_f^3))\,,\\
   D_{\text{S}}=&-26.4435 x^3 + (-1521.21 - 218.159 L + (49.0568 + 13.2218 L) n_f]\,,
   \label{eq:adler0}
\end{align}
where $L=\log(\mu^2/q^2)$ and $x=\alpha_s(\mu)/\pi$. The total contribution to the Adler function is given by:
 \begin{align}
     D^{em}=\sum_{f}(e_f^2) D_{\text{NS}}+(\sum_{f}e_f)^2D_{\text{S}}
     \label{eq:adlerEM}
 \end{align}
where $e_f$ is the charge of the $n_f-$flavored quark and the singlet contribution vanishes for the $n_f=3$ case. These expressions are relevant for dimension zero contribution to the $e^+e^-\rightarrow$ hadrons, hadronic Z- and $\tau$ decays, electromagnetic contributions to muon $\left(g-2\right)_{\mu}$, $J/\psi$ and $\Upsilon$ systems. \par
Predictions for the higher-order terms using large$-\beta_0$ approximation and Borel models for hadronic $\tau$ decay width in Ref.~\cite{Beneke:2008ad}:
\begin{align}
    \delta D^{em}\vert_{n_f=3}=&283 x^5 + 3275 x^6 + 18800 x^7 + 388000 x^8 + 919000 x^9 +
    8.37\times10^7 x^{10}\nonumber\\& - 5.19\times10^8 x^{11} + 3.38\times10^{10} x^{12}\,,
    \label{eq:AdlerTau}
\end{align}
Other predictions using D-log Pad\'e approximants for hadronic $\tau$ decay to $\ordas{6}$ can be found in Ref.~\cite{Boito:2018rwt}.
\subsection{Leading mass correction to \texorpdfstring{$\text{R}-$}{}ratio}
\label{app:Rem_mass_correction}
The massless and leading-order massive corrections to $e^+e^-\rightarrow\text{hadrons}$ are now known to $\ordas{4}$ and  can be found in Ref.~\cite{Chetyrkin:1990kr,Chetyrkin:1994ex,Chetyrkin:2000zk,Baikov:2004ku}. The leading mass correction coefficient of the Adler function for the vector current correlator for $n_f=3$ case in FOPT reads:
\begin{align}
	D^{(2)}_{V}=&12 x +x^2 [113.5 + 51 L] +x^3 [1275.89 + 876.75 L + 
	165.75 L^2]+x^4[16496.5 \nonumber\\&\bs+ 13351.6 L + 4233.16 L^2 + 
	483.438 L^3] 
\end{align}
 and the corresponding expression in RGSPT reads:
 \begin{align}
 	D^{(2)}_{V}=&12 x\hs w^{-17/9} +x^2 \left[(134.981 - 40.2963 L_w)w^{-26/9}  - 21.4815w^{-17/9}\right] + x^3 \big[(
 	1551.31 \nonumber\\&- 764.876 L_w + 103.477 L_w^2)w^{-35/9} - (271.34 - 72.135 L_w)w^{-26/9}  - 4.073w^{-17/9}\big]\nonumber\\&+ x^4 \big[w^{-44/9}(
 	19944.3 - 12084.9 L_w + 2827.98 L_w^2 - 238.465 L_w^3) + w^{-35/9} (-3340.89\nonumber\\& + 1521.79 L_w - 185.236 L_w^2) + (-114.043 + 13.679 L_w)w^{-26/9}  + 7.12168w^{-17/9}\big]
 \end{align}

\section{Adler function for the scalar current correlator}\label{app:D_scalar}
The Adler function of the scalar current is known to $\ordas{4}$ \cite{Becchi:1980vz,Broadhurst:1981jk,Chetyrkin:1996sr,Baikov:2005rw,Gorishnii:1990zu,Gorishnii:1991zr,Herzog:2017dtz} relevant for the process $\text{H}\rightarrow\overline{\text{b}}\text{b}$, and leading mass correction for the hadronic $\tau$ decay for the longitudinal contributions. Its numerical value for $n_f$ quark flavor is given by:
\begin{align}
    D_2= &1 + x (5.667 + 2 L) + x^2 (51.57 + 35.33 L +
    4.75 L^2 + (-1.907 - 1.22 L -
    0.167 L^2)  n_f)  \nonumber\\&+ x^3 (648.7 + 509.6 L + 147.29 L^2 +
    11.88 L^3 - (63.74 + 42.12 L + 11.54 L^2 +
    0.94 L^3)  n_f \nonumber\\&+ (0.929 + 0.582 L + 0.204 L^2 +
    0.0185 L^3)  n_f^2) +x^4 (9470.76 + 8286.31 L +
    3047.01 L^2 \nonumber\\&+ 536.76 L^3 +
    30.4297 L^4 - (1454.28 + 1202.9 L + 398.02 L^2 + 68.11 L^3+ 3.91 L^4)  n_f \nonumber\\&+ (54.783 + 45.723 L + 14.688 L^2 +
    2.723 L^3 + 0.166 L^4)  n_f^2 + (-0.454 - 0.453 L \nonumber\\&-
    0.145 L^2 - 0.034 L^3 - 0.0023 L^4)  n_f^3)+\ordas{5}\,,
\end{align}
where $x=\as(\mu^2)/\pi$ and $L=\log(\mu^2/q^2)$

	\section{Wilson coefficients \texorpdfstring{$C_1$}{} and \texorpdfstring{$C_2$}{}}\label{app:Wilson}
   The Wilson coefficients needed for the hadronic Higgs decays, $C_1$ and $C_2$ are known to $\order{\alpha_s^5}$ and are obtained using the four loops decoupling relations~\cite{Liu:2015fxa,Schroder:2005hy,Chetyrkin:2005ia} and expressions can be found in Refs.~\cite{Herzog:2017dtz,Gerlach:2018hen}. Their value in the different mass schemes are calculated using the quark mass relations \cite{Gray:1990yh,Chetyrkin:1999ys,Chetyrkin:1999qi,Melnikov:2000qh,Marquard:2015qpa,Marquard:2016dcn} and their numerical values in the $\msbar$-, SI-  and OS- schemes are given by:
    \begin{align}
        C_1^{\msbar}&=\frac{-x}{12}\bigg[1 + 2.750 x + x^2[9.642-  0.698 n_l + L_m(1.188 + 0.333 n_l) ]+x^3 [47.370 - 7.69n_l  \nonumber\\&\bs- 0.221n_l^2+   L_m^2 (3.266 + 0.719n_l- 0.056n_l^2) +
        L_m(6.017 + 1.019n_l+ 0.045n_l^2)]  \nonumber\\& \bs+x^4 [311.780 - 62.368n_l+  1.616n_l^2- 0.034n_l^3 +   L_m^2 (21.878 + 1.796n_l\nonumber\\&\hspace{1.8cm}- 0.092n_l^2 - 0.011n_l^3) +   L_m^3 (8.980 + 1.432n_l- 0.273n_l^2 + 0.009n_l^3)\nonumber\\&\hspace{1.8cm} +    L_m(26.504 - 10.211n_l- 2.426n_l^2 + 0.093n_l^3)] \bigg]\\
        C_1^{SI}&=C_1^{\msbar}+x^4 L_m(-2.375 - 0.667 n_l) + x^5 L_m[-21.700 - 4.420n_l+ 0.003n_l^2 \nonumber\\&\bs\bs+
        L_m(-16.526 - 3.649n_l+ 0.278n_l^2)]\\
        C_1^{OS}&=C_1^{\msbar}+x^4[-3.167- 0.889 n_l + L_m(-2.375 - 0.667 n_l) ]\nonumber\\& \hspace{1.3cm}+ x^5[-52.197 -
        10.390n_l+ 0.575n_l^2 -
        L_m(47.825 + 10.170n_l- 0.448n_l^2)\nonumber\\&\hspace{1.3cm}+ L_m^2 (-16.526 - 3.649n_l+ 0.278n_l^2)]\,,
        \end{align}
        \begin{align}
        C_2^{\msbar}&=1 + x^2[0.278 - 0.333 L_m]+ x^3[2.243 - 1.528 L_m- 0.917 L_m^2 + (0.245 + 0.056 L_m^2) n_l] \nonumber\\& \hspace{.6cm}+ x^4 [2.095 - 2.627 L_m- 6.736 L_m^2 - 2.521 L_m^3 + (-0.01 - 0.096 L_m-
        0.009 L_m^3)n_l^2\nonumber\\&\hspace{.6cm}+ (0.31 + 3.142 L_m+ 0.479 L_m^2 +  0.306 L_m^3)n_l]+x^5 [65.144 + 13.373n_l- 3.642n_l^2 \nonumber\\&\hspace{.6cm}+ 0.076n_l^3 + L_m(-121.028 + 27.852n_l- 0.959n_l^2 - 0.001n_l^3)+ L_m^2 (-17.373\nonumber\\&\hspace{.6cm} + 19.162n_l- 1.581n_l^2 + 0.032n_l^3)+ L_m^3 (-25.654 + 3.630 n_l - 0.126n_l^2)  \nonumber\\&\hspace{.6cm}+ L_m^4 (-6.932 + 1.260n_l- 0.076n_l^2 + 0.002n_l^3) ]\\
        C_2^{SI}&= C_2^{\msbar}+x^3 0.667 L_m +x^4 [L_m^2 (4.639 - 0.278 n_l) +
        L_m(5.769 - 0.093 n_l)]  \nonumber\\&\hspace{1.3cm}+ x^5[
        L_m(29.056 - 8.266n_l+ 0.173n_l^2)+L_m^2 (60.290 - 4.793n_l+  0.046n_l^2) \nonumber\\&\hspace{1.3cm}+ L_m^3 (22.261 - 2.673n_l+ 0.080n_l^2)]\\
        C_2^{OS}&=C_2^{\msbar}+x^3[0.889 + 0.667 L_m] + x^4[14.222 - 0.694 n_l + L_m(13.102 - 0.537 n_l) \nonumber\\ &\hspace{1.3cm}+ L_m^2 (4.639 - 0.278 n_l)]+ x^5[206.029 - 30.829n_l+ 0.690n_l^2 \nonumber\\&\hspace{1.3cm}+L_m(203.548 - 27.199n_l+ 0.636n_l^2) +  L_m^2 (100.623 - 9.682n_l+ 0.194n_l^2)\nonumber\\&\hspace{1.3cm}+ L_m^3 (22.261 - 2.673n_l+ 0.080n_l^2) ]\,.
    \end{align}
where $L_m=\log(\mu^2/m_q^2)$ and $m_q$ are the mass of the quark in the mentioned quark mass scheme.\par
 
\section{Adler Functions Relevant for the Hadronic Higgs Decay Width.} \label{app:D_higgs}
The Adler functions relevant to the hadronic Higgs decay to compute the $\Delta_{i,j}$ are presented in this section. The $\Delta_{ij}$ are obtained from the discontinuity of the $D_{ij}$ with appropriate factors.
\subsection{FOPT expressions}
\begin{align}
    D_{11}=&1 + x[12.417 + 3.833 L_H]  + x^2[104.905 + 81.063 L_H + 
    11.0208 L_H^2] \nonumber\\& +x^3 [886.037 + 971.268 L_H + 333.899 L_H^2 + 
    28.1644 L_H^3]\nonumber\\&+ x^4(8723.76 + 10408.8 L_H + 5279.62 L_H^2 + 
    1119.89 L_H^3 + 67.4771 L_H^4)\nonumber\\&+\frac{m_b^2}{M_H^2}\bigg[6 x + x^2[202.046 + 69.5 L_H + 3 L_H^2]  + x^3[4069.51 + 2622.94 L_H  \nonumber\\&\bs\bs+ 
    462.167 L_H^2+ 17.5 L_H^3]\bigg]\,,\\
    D_{12}=&-x[30.67 + 8 L_H] - x^2[524.701 +280.44 L_H + 
    31.33 L_H^2]\nonumber\\&- x^3[7093.07 + 5337.21 L_H +1316.94 L_H^2 + 
    91.39 L_H^3]\,.
\end{align}
The analytic results for $\Delta_{11}$ can be obtained from Ref.~\cite{Davies:2017xsp,Herzog:2017dtz,Davies:2017rle}.
\subsection{RGSPT expressions}
\label{app:D11_RGSPT}
\begin{align}
D_{1,1}^{\Sigma}  =&w^{-2} + x(14.9384w^{-3} - 2.52174 L_w w^{-3} - 2.52174w^{-2})  + x^2(140.526
w^{-4}  \nonumber\\&- 59.6857 L_w w^{-4}+ 4.76938 L_w^2 w^{-4} - 37.441 w^{-3} + 
6.360 L_w w^{-3} + 1.820w^{-2})  \nonumber\\&+ x^3(1233.01 w^{-5} - 783.997 L_w
w^{-5} + 156.525 L_w^2 w^{-5} - 8.018 L_w^3 w^{-5} - 348.927w^{-4}  \nonumber\\&+ 
149.642 L_w w^{-4} - 12.027 L_w^2 w^{-4} + 18.484w^{-3} - 4.589L_w
w^{-3} - 16.533w^{-2}) \nonumber\\& + x^4(11878.4w^{-6} - 8761.86 L_w w^{-6} +
2668.65 L_w^2w^{-6} - 339.04 L_w^3 w^{-6} + 49.6729 w^{-2}  \nonumber\\&+ 12.6372 L_w^4w^{-6} - 
3037.83w^{-5} + 1948.48 L_w w^5 - 392.522 L_w^2w^5 + 
20.2195 L_w^3w^{-5} \nonumber\\&+ 49.8197w^{-4} - 75.7019 L_w w^{-4} + 
8.67931 L_w^2 w^{-4}- 216.349w^{-3} + 41.6915 L_w w^{-3}) \nonumber\\&+\frac{m_b^2}{M_H^2w^{24/23}}\bigg[(0.8166 + 0.8166w^{-2} - 1.633 w^{-1} + x(1.7599 + 30.287w^{-3}\nonumber\\&  - 
3.134 L_w w^{-3}- 34.8133 w^{-2} + 4.20822 L_ww^{-2} + 8.76682 w^{-1} - 
1.07444 L_w w^{-1} )\nonumber\\&  + x^2[1.499 + 638.823 w^{-4} - 158.362 L_ww^{-4} + 
7.988 L_w^2 w^{-4} - 551.269 w^{-3} \nonumber\\&+ 138.9 L_w w^{-3} - 8.074 L_w^2
w^{-3}+ 97.462w^{-2} - 23.943 L_w w^{-2} + 1.384 L_w^2w^{-2}\nonumber\\&  + 
15.531w^{-1} - 2.315 L_w w^{-1}]  + x^3[-12.839 + 10698.6 w^{-6}- 
4262.05 L_w w^{-5} \nonumber\\&+ 513.596 L_w^2 w^{-5}  - 16.93 L_w^3 w^{-5} - 
7966.85w^{-4} + 2985.67 L_w w^{-4} - 364.26 L_w^2 w^{-4}\nonumber\\& + 
13.7218 L_w^3 w^{-4} + 1189.89w^{-3} - 404.192 L_w w^{-3} + 
47.685 L_w^2 w^{-3} - 1.77 L_w^3 w^{-3}\nonumber\\& + 129.34 w^{-2} - 42.9368 L_w
w^{-2} + 2.98299 L_w^2 w^{-2} + 31.3499 w^{-1} - 1.97163 L_w w^{-1}]\bigg]
\end{align}
\begin{align}
    D_{12}^{\Sigma}=w^{-24/23}&\bigg\lbrace (4.17391(1-w^{-1})  + x [-0.409 + (-78.7983 + 10.7543 L_w)w^{-2}  \nonumber\\&+ (48.5405 - 5.49158 L_w)
    w^{-1})] + x^2[-4.37668 + w^{-3}(-1032.38 \nonumber\\&+ 315.943 L_w  - 20.6345 L_w^2)
     + (532.464 - 131.992 L_w + 7.07469 L_w^2)
    w^{-2} \nonumber\\&+ (-20.4045 + 0.53794 L_w)
    w^{-1}]  +x^3 \big[-29.29 + (-12441.7 + 5661.76 L_w  \nonumber\\&- 831.402 L_w^2 + 
    35.067 L_w^3)w^{-4} + (
    5715.82 - 2209.72 L_w + 262.174 L_w^2 \nonumber\\&- 9.04953 L_w^3)
    w^{-3} + (-310.481 + 53.2516 L_w - 0.693018 L_w^2)
    w^{-2} \nonumber\\&+ (-27.3994 + 5.75835 L_w)w^{-1}\big] \bigg\rbrace
\end{align}
\begin{align}
    D_{22}^{\Sigma}=w^{-24/23}& \bigg\lbrace 1 + x[-2.35098 + (8.01764 - 1.31569 L_w)w^{-1}] +x^2 \big[1.144 + (
    59.6174 \nonumber\\&- 22.3168 L_w + 1.69498 L_w^2)
    w^{-2} + (-18.7293 + 3.09316 L_w)w^{-1}\big]\nonumber\\&+ x^3\big[3.74239 + (
    482.958 - 256.916 L_w + 44.9568 L_w^2 - 2.16812 L_w^3)
    w^{-3}\nonumber\\& + (-138.124 + 52.1572 L_w - 3.98485 L_w^2)w^{-2} + (
    4.65204 - 1.50502 L_w)w^{-1}\big]  \nonumber\\&+ x^4\Big[-10.6936+ (
    45.18 - 4.92 L_w)w^{-1} - (8.45 + 13.88 L_w - 1.94 L_w^2)w^{-2}\nonumber\\&+ (
    4598.16 - 2786.2 L_w + 711.6 L_w^2 - 79.135 L_w^3 + 
    2.763 L_w^4)
    w^{-4} \nonumber\\&+ (-1111.99 + 595.803 L_w - 105.099 L_w^2 + 5.09719 L_w^3)
    w^{-3} \Big] \bigg\rbrace
\end{align}

\end{document}